\documentclass[aoas]{imsart}

\RequirePackage{amsthm,amsmath,amsfonts,amssymb}
\RequirePackage[authoryear]{natbib}
\RequirePackage[colorlinks,citecolor=blue,urlcolor=blue]{hyperref}
\RequirePackage{graphicx}
\usepackage{tabularray}
\usepackage{booktabs}
\startlocaldefs
\newcommand{\iid}{\overset{\scriptscriptstyle\smash{\mathrm{iid}}}{\sim}}
\providecommand{\tightlist}{\setlength{\itemsep}{0pt}\setlength{\parskip}{0pt}}

\endlocaldefs

\begin{document}

\begin{frontmatter}
\title{How long should a block be?}
\runtitle{How long should a block be?}

\begin{aug}
\author[A]{\fnms{Léo R.}~\snm{Belzile}\ead[label=e1]{leo.belzile@hec.ca}\orcid{0000-0002-9135-014X}}
\author[B]{\fnms{Anthony C.}~\snm{Davison}\ead[label=e2]{anthony.davison@epfl.ch}\orcid{0000-0002-8537-6191}}
\address[A]{Department of Decision Sciences, HEC Montréal\printead[presep={,\ }]{e1}}

\address[B]{Institute of Mathematics, École polytechnique fédérale de Lausanne\printead[presep={,\ }]{e2}}
\end{aug}

\begin{abstract}
 The block maximum method, which is widely used in extreme value analysis, uses a generalized extreme value distribution to approximate that of the maximum of $m$ observations.  The quality of this approximation depends on the value of $m$ and may be poor if $m$ is too small.  Surprisingly  little attention has been paid to the choice of the block length, although a good choice is crucial to the success of the method.  In this paper we assess the effect of taking excessively long blocks in terms of asymptotic relative efficiency, and propose likelihood-based approaches and graphical diagnostics to determine whether a proposed block length is suitable, allowing for potential rounding and left-censoring of observations. We investigate our ideas using simulation and illustrate them using wind speed, river flow and rainfall data.
\end{abstract}

\begin{keyword}
\kwd{Asymptotic relative efficiency}
\kwd{Block maximum method}
\kwd{Generalized extreme value distribution}
\kwd{Graphical diagnostic}
\kwd{Likelihood ratio test}
\kwd{Penultimate approximation}
\end{keyword}

\end{frontmatter}


\section{Introduction}\label{introduction}

The block maximum method is the oldest and best-established approach to
inference for sample extremes. The idea is to split a time series of
background data into equal-length disjoint blocks of successive
observations, and then to fit the generalized extreme value distribution
to the block maxima. This fitted distribution is then used to
extrapolate further into the tail of the data, typically with the goals
of estimating either the probability that some high level will be
exceeded or the upper quantiles of the distribution of the background
data. This approach to risk analysis was promoted for engineering
applications by Emil Gumbel, whose ideas were summarised in
\citet{Gumbel:1958}, and is based on the extremal types theorem,
which shows that under mild assumptions the generalized extreme value
distribution is the only possible limit for the distribution of
linearly-rescaled block maxima --- see, e.g., \citet{Coles:2001} or
\citet{Beirlant:2004}. The block maximum method can also be applied
to block minima, simply by analysing the block maxima of the negated
data, but as is common in the literature we consider maxima throughout.

Another approach to the analysis of extremes, which involves fitting the generalized Pareto distribution to exceedances of a high threshold, has its roots in hydrology \citep{Todorovic.Zelenhasic:1970}, and is justified mathematicallly by the Balkema--de~Haan--Pickands theorem \citep{Balkema.deHaan:1974, Pickands:1975}, which establishes that the only nondegenerate limiting distribution of rescaled threshold exceedances is generalized Pareto.  This method has been used in a vast range of applications since the publication of \citet{Davison.Smith:1990}.

In addition to their theoretical support, the generalized extreme-value
and generalized Pareto distributions are often found to fit maxima and
exceedances well. They satisfy natural stability properties that provide
a principled approach to extrapolation outside the available data. From
an empirical viewpoint, therefore, it is important to verify that the
stability properties apply in finite samples, by checking
that the block length or threshold are large enough. Theshold selection has been widely considered --- see
\citet{Belzile.Davison:2026} for a recent review --- but formal
consideration of the choice of block length has been limited.

This choice leads to a bias-variance trade-off
\citep[e.g.,][Section 3.3.1]{Coles:2001}. Blocks that are too short
lead to biased extrapolations and potentially poor fit of the limiting
generalized extreme value model, whereas blocks that are too long lead
to improved approximations, but smaller numbers of maxima and more
variable estimators. One can consider the asymptotic properties of estimators based on block maxima under domain of attraction
assumptions --- that is, allowing for the fact that the generalized
extreme value distribution is only valid asymptotically --- when
contrasted with use of threshold exceedances
\citep{Bucher.Zhou:2021}. Such results are of theoretical interest
but limited practical value, as the framework requires knowledge of the
unknown distribution of the data and the bias of the estimators cannot
be computed in finite samples.

Most, if not all, work on the choice of block length suggests
determining \(m\) through goodness-of-fit measures such as
Anderson--Darling or Kolmogorov--Smirnov tests; \citet{Wang:2016}
suggest pooling results of multiple tests. In unpublished work,
\citet{Dkengne:2020} suggest using max-stability to derive parameters
that should be constant as a function of \(m\), and building diagnostic
plots that allow the selection of a value of \(m\) above which such
estimates stabilize, in the same vein as threshold stability plots
\citep{Davison.Smith:1990,Coles:2001}. Such diagnostics can suffer
from issues of multiple testing and are strongly correlated due to
sample overlap. One can also consider the stability of estimates of the
extremal index from block maxima as the block length increases
\citep{Northrop:2015,Berghaus.Bucher:2018}. There is no satisfactory
formal approach to choosing the block length, and the purpose of this
article is to fill this gap.

Pragmatic considerations are crucial in applications. If the background
data are available and are reasonably stationary, then exceedance
analysis is often employed, but any clustering of extremes must then be
handled, and it may be preferable to dodge this. In some cases,
historical maxima and minima may only be available annually, so
exceedance analysis cannot be performed and the shortest block is of
length one year. More recent data may be recorded at higher frequencies,
so shorter blocks could in principle be used. Taking monthly blocks,
say, then increases the number of maxima by a factor of 12, but the
resulting gain in information may be offset by having to allow for
seasonality in the monthly maxima by some form of regression analysis.
Despite this many practitioners opt for monthly or yearly blocks, owing
to their ready interpretation. The increased availability of high
frequency environmental data permits use of much shorter blocks, which
poses the question of whether these are suitable for extreme value
analysis or whether further aggregation is required.

\subsection{Motivating example} \label{sec-motivating-example}

To motivate the need for a formal testing procedure for determining the block length, we consider daily data from the
\href{https://raws.dri.edu/cgi-bin/rawMAIN.pl?caCCHB}{Western Regional
Climate Center} for the
\href{https://www.ncdc.noaa.gov/cdo-web/datasets/GHCND/stations/GHCND:USR0000CCHB/detail}{Cheeseboro}
weather station in Southern California; these are discussed in more detail in Section~\ref{application-1-hourly-maximum-wind-speed-at-cheeseboro}. The data contains 921 observations of daily maximum gust wind speed (in miles per hour,
mph) for 1 January to 1 February inclusive between 1996 and 2026. Measurements are recorded to the
nearest mile per hour, with a sample maximum of 92mph.

\begin{figure}[ht!]
\centering
\includegraphics[width=\textwidth]{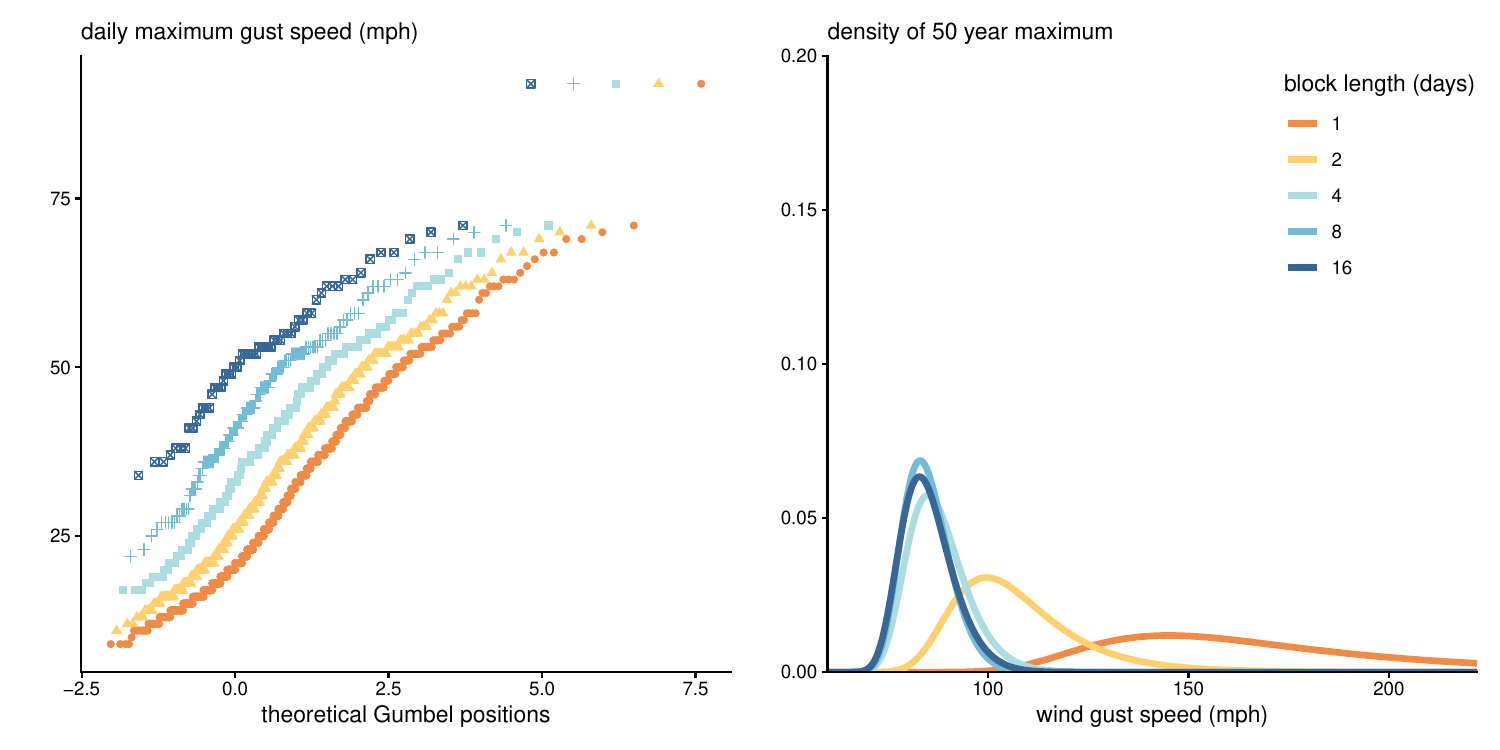}
\caption{\label{fig-density-50year}Cheeseboro wind gust data: Gumbel
plot of \(m\)-day maxima against standard Gumbel plotting positions
\(-\log[-\log\{i/(n_m+1)\}]\) (left) and the estimated GEV density of
the 50-year maximum for the Cheeseboro wind gust data evaluated at the
maximum likelihood estimators, for different block lengths (right).}
\end{figure}

Figure~\ref{fig-density-50year} shows a Gumbel quantile-quantile plot of the data for different block length $m$, in the spirit of \citet{Cox.Isham.Northrop:2002}. If the model for small $m$ was adequate, we would expect to see parallel lines above for larger values of $m$. Such plot is difficult to interpret because of the sample overlap of the maxima for different values of $m$, but there is noticeable warping for $m\in \{1,2\}$. The right panel shows the estimated densities of 50-year maximum for different block lengths $m$: the densities for \(m=1\) and \(m=2\) are much more dispersed than are the others. The curves suggest wind gust speeds very much beyond the observed maximum of 92mph for $m=1$, close to the median for $m=2$, whereas for longer block lengths the curves more or less agree and suggest more moderate extremes. The figure shows that taking insufficiently long blocks could greatly impact risk assessment.

\section{Extreme value theory}\label{extreme-value-theory}
\subsection{Extremal types theorem}\label{extremal-types-theorem}

The extremal types theorem
\citep{Fisher.Tippett:1928,vonMises:1936,Gnedenko:1943} states that,
if there exist scale and location sequences \(c_m>0\) and
\(d_m \in \mathbb{R}\) such the distribution of the maximum of \(m\)
independent and identically distributed variables with distribution
function \(G\) converges to a nondegenerate distribution \(F\) after
affine renormalization, i.e., \(\lim_{m\to \infty}G^m(c_mx+d_m)=F(x)\)
for all real \(x\), then \(F\) must be a generalized extreme value (GEV)
distribution, \begin{align*}
F(x; \mu, \sigma, \xi)=
\begin{cases}\exp\left[-\left\{1+\xi (x-\mu)/\sigma\right\}_{+}^{-1/\xi}\right], & \xi \neq 0, \\
\exp\left[-\exp\left\{-(x-\mu)/\sigma\right\}\right], & \xi =0.
\end{cases}
\end{align*} The parameters \(\mu\) and \(\sigma\) are the location and
scale parameters of the distribution; the shape parameter \(\xi\)
determines its support and the weight of its upper tail, and thus plays
a crucial role in extrapolation to higher values.

The extremal types theorem extends to stationary time series: under a
condition \(D(u_n)\) \citep{Leadbetter:1983} that limits long range
dependence of extremes, there exists an extremal index
\(\theta \in (0,1]\) for which\\
\[
G^{m}(c_{m\theta}x+d_{m\theta}) = G^{m\theta}(c_mx+d_m) \to F(x), \quad m\to\infty,
\] where \(c_{\theta m} = c_m\theta^\xi\) and
\(d_{\theta m}=d_{m}+c_m (\theta^\xi-1)/\xi\).

This limit arises when the block length goes to infinity, but the result
is applied for finite \(m\): if \(n\) years of daily data are available,
then, among other possibilities, the analyst might fit the GEV
distribution to the maxima of \(12n\) monthly blocks, or \(4n\) seasonal
blocks or \(n\) annual blocks. The unknown parameters are estimated from
these maxima, and clearly the precision of the resulting estimates will
increase with the number of blocks, while the quality of the fit will
improve with the block length. What length should we choose?

A key issue, the speed at which \(G^m\) converges to \(F\), depends on
the upper tail properties of \(F\), which determine its pre-limit, or
penultimate, behaviour. This can be studied under mild continuity
conditions \citep[cf.][]{Smith:1987}. Suppose that \(G\) is twice
differentiable with density \(g\) and possibly infinite upper endpoint
\(x^{\star}\) to the support of \(g\), and let
\(\phi(x)=-G(x)\log G(x)/g(x)\). Then, for the convergence of \(G^m\) to
the GEV distribution, it is sufficient that
\(\lim_{x \to x^{\star}} \phi'(x)=\xi_0 \in \mathbb{R}\). In practice, due to finite block lengths and model
misspecification, this limiting approximation may be improved by using a
different value of \(\xi\); indeed, the expectations of the maximum
likelihood estimators for the shape vary slowly with the block length
\(m\). \citet{Smith:1987} shows that the penultimate GEV
approximation obtained by taking \(F\) with location \(d_m\), scale
\(c_m\) and shape \(\xi=\phi'(d_m)\) leads to a smaller Hellinger
distance between it and the true distribution \(G^m\) than does taking
the limiting, constant shape parameter, \(\xi_0\). We appeal to these
results to build realistic alternatives in our simulation studies.

\subsection{Max-stability}\label{max-stability}

The GEV is characterized by \emph{max-stability}, whereby the maximum of
a random sample of observations can be linearly rescaled to have the
same distribution as an original observation. Let
\(X_{(1)} \leq \cdots  \le X_{(m)}\) denote the \(m\) order statistics
of a random sample
\(X_1,\ldots, X_m \iid\mathsf{GEV}(\mu, \sigma, \xi)\). Their maximum
\(Y=X_{(m)} = \max(X_1,\ldots, X_m)\) has distribution function
\(F_{(m)}(x)=F^m(x)\) and probability density function
\begin{equation}\protect\phantomsection\label{eq-maxstable}{
f_{(m)}(x) = m F(x; \mu, \sigma, \xi)^{m-1}f(x; \mu, \sigma, \xi) = f(x; \mu_m, \sigma_m, \xi),
}\end{equation} where \(f\) and \(F\) are the GEV density and
distribution functions, and where
\begin{equation}\protect\phantomsection\label{eq-mum-eq}{
\mu_m = \begin{cases} \mu + \sigma (m^\xi-1)/\xi, &\xi\neq 0,\\ \mu+\sigma\log m, &\xi=0,\end{cases}\qquad
\sigma_m = \sigma m^\xi,
}\end{equation} Max-stability implies that the distributions of \(Y\)
and the \(X\)'s are of the same form, with the same shape parameter and
with a known deterministic relation between their respective location
and scale parameters. This is the basis of the development below, since
it suggests that stability can be assessed by testing whether the
relationships in Equation~\ref{eq-mum-eq} between the distributions of
the \(X\)s and \(Y\) hold.

\subsection{Information loss}\label{information-loss}

Before constructing tests for max-stability, it is natural to ask
whether replacing a sample by its maximum leads to a large loss of
information about the underlying distribution, in light of eq.~\eqref{eq-maxstable}. If not, then taking
blocks slightly larger than strictly necessary would lead to robust
inferences without much loss of precision, suggesting that larger blocks
should be used when in doubt. As we shall see below, the reduction of
precision depends on the target of inference.

To quantify the loss of information due to the analysis of block maxima,
we argue as follows. The \(3\times 3\) Fisher information matrix for a
single \(\mathsf{GEV}(\mu, \sigma, \xi)\) observation may be written as
\(\mathcal{I}(\xi) = D^{-1} K(\xi) D^{-1}\), where
\(D=\mathrm{diag}(\sigma,\sigma,1)\) and \(K(\xi)\) is the Fisher
information for the \(\mathsf{GEV}(0, 1, \xi)\) distribution
\citep{Prescott.Walden:1980}. Thus, the variance matrix for the
maximum likelihood estimator based on a sample of \(m\) such
observations is \(m^{-1} D K(\xi)^{-1} D\). The Fisher information
matrix based on the sample maximum \(X_{(m)}\), which has distribution
\(\mathsf{GEV}(\mu_m, \sigma_m, \xi)\), is \(D_m^{-1}K(\xi)D_m^{-1}\),
where \(D_m=\mathrm{diag}(\sigma_m,\sigma_m,1)\). However this matrix
corresponds to the parameters
\({\boldsymbol{\vartheta}}_m=(\mu_m,\sigma_m,\xi)^\top\), whereas that
for the random sample relates to the parameters
\({\boldsymbol{\vartheta}}=(\mu,\sigma,\xi)^\top\) of the original
model. Thus the Fisher information matrix for
\({\boldsymbol{\vartheta}}\) based on the sample maximum is \[
\mathcal{I}_m(\xi)= \frac{\partial {\boldsymbol{\vartheta}}^\top_m}{\partial {\boldsymbol{\vartheta}}}D_m^{-1}K(\xi)D_m^{-1}
\frac{\partial {\boldsymbol{\vartheta}}_m}{\partial {\boldsymbol{\vartheta}}^\top}.
\] The overall asymptotic relative efficiency of basing inference on the
maximum likelihood estimator \(\widehat{\boldsymbol{\vartheta}}_m\) that
uses only the sample maximum rather than tthe entire sample,
\(\widehat{\boldsymbol{\vartheta}}\), is \[
\frac{|\mathsf{Var}(\widehat{\boldsymbol{\vartheta}})|^{1/3}}
{|\mathsf{Var}(\widehat {\boldsymbol{\vartheta}}_m)|^{1/3}}  = \frac{|\mathcal{I}_m(\xi)|^{1/3}}
{|m\mathcal{I}(\xi)|^{1/3}}  = \frac{1}{m^{1+2\xi/3}},
\] so at the Gumbel model, i.e., when \(\xi=0\), the overall loss of
information, \(1/m\), corresponds to replacing a sample of \(m\)
independent Gumbel variables by a single such variable. The rate of this
overall loss increases as \(\xi\) increases and decreases as \(\xi\)
decreases, with a limit of \(m^{-2/3}\) when \(\xi\to-1/2\), which is
the lower limit for regularity of maximum likelihood estimation. When
\(\xi<-1/2\), the sample maximum is super-efficient, in the sense that
it converges to the upper support point of the density more rapidly than
in a regular case. The quality of dependence on \(\xi\) seems plausible:
as \(\xi\) increases, information about the upper tail of the
distribution is increasingly spread among the sample order statistics,
so the reduction from losing all but \(X_{(m)}\) grows, whereas as
\(\xi\) decreases, \(X_{(m)}\) becomes closer to super-efficient,
and the relative loss due to dropping the rest of the sample declines. A
related phenomenon appears when generalized Pareto samples are right-censored: when \(\xi<0\), the
asymptotic efficiency relative to estimation from a full sample can drop
to 50\% when the top 1\% of a sample is censored
\citep[Table~1]{Davison.Smith:1990}.

\begin{figure}[tbp!]

\centering{

\includegraphics[width=\textwidth]{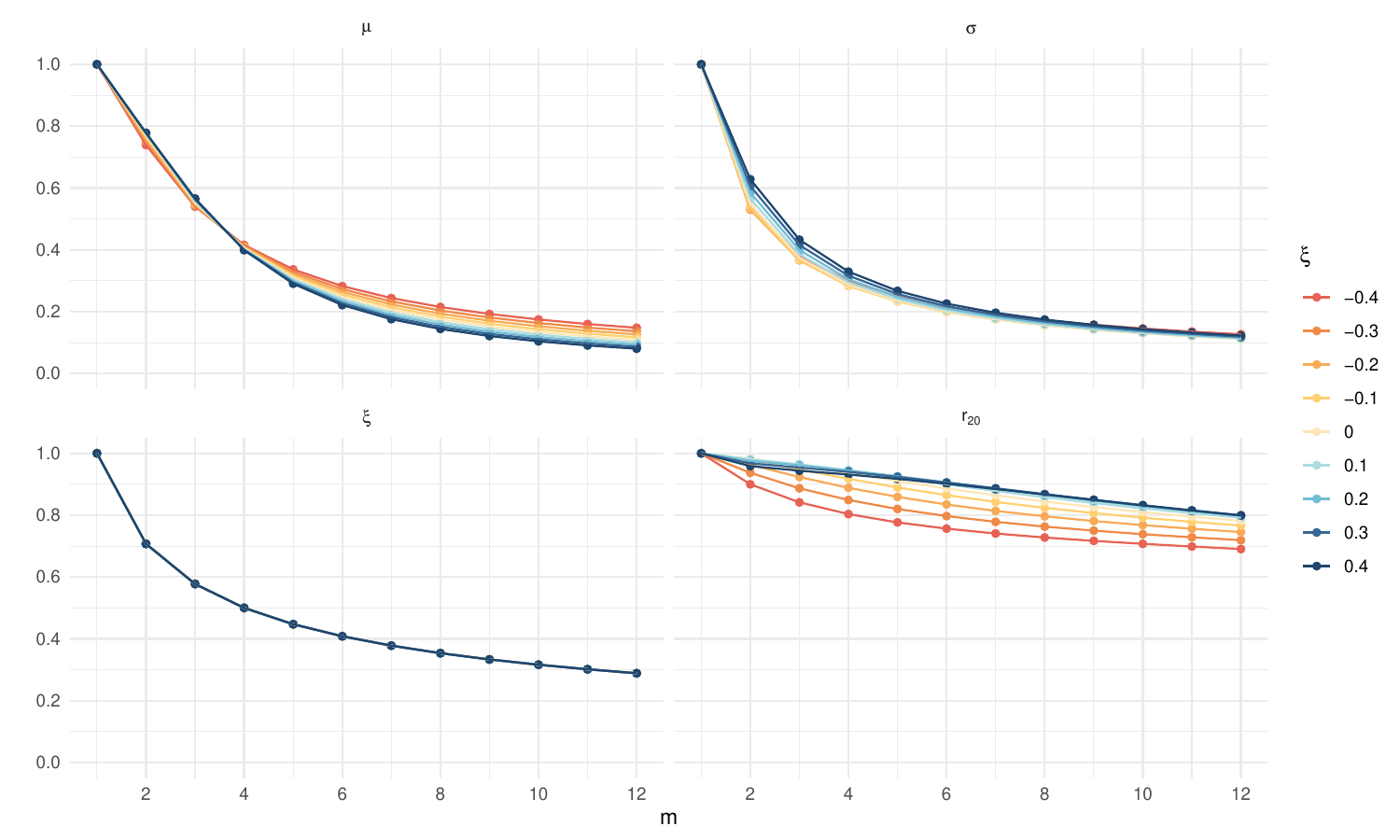}

}

\caption{\label{fig-ARE}Ratios of lengths of asymptotic confidence
intervals for maximum likelihood estimators of \(\mu\), \(\sigma\),
\(\xi\) and the 20-observation return level \(r_{20}\), based on a
random sample of \(m\) GEV observations and on the sample maximum only.}

\end{figure}

The efficiencies for estimation of individual parameters
may be computed as the ratios of the diagonal elements of the variance
matrices. Figure~\ref{fig-ARE} shows the square roots of these ratios
for different values of \(\xi\) and for \(m\) in the interval
\([1,12]\), and the corresponding ratio for estimation of the
20-observation return level. These square roots have a direct
interpretation as the ratio of the asymptotic lengths of confidence
intervals based on the full sample rather than on \(X_{(m)}\) alone.
Thus, for example, the upper right-hand panel shows that intervals for
the scale parameter \(\sigma\) based on the entire sample are
respectively roughly 0.6 and 0.12 times shorter than those based on
maxima of samples of sizes \(m=2\) and \(m=12\).

The ratios for the parameter estimators drop rapidly with \(m\) but
depend little on \(\xi\); indeed, the ratio for \(\xi\) depends on \(m\)
but not on \(\xi\) itself. The ratio for the 20-observation return
level, or equivalently the 0.95 quantile, \(r_{20}\), drops more slowly
that those for the individual estimators, taking values between 0.7 and
0.8 when \(m=12\), with steeper initial drops for negative \(\xi\). It
seems that estimation of the parameters, particularly the location and
scale, degrades rapidly when maxima are used in preference to the full
sample, but this drop is much less dramatic for upper quantiles.

The supplementary material contains the corresponding ratios for certain
other return levels, which show the same broad pattern as in
Figure~\ref{fig-ARE}. The contrast between the ratios for estimation of
the parameters and of the return level might seem puzzling, but
estimation of the parameters of the original GEV distribution
essentially involves extrapolation from maxima back to a lower level,
which increases the uncertainty, whereas a moderate-level quantile will
typically lie within the span of a sample of maxima, and thus be better
estimated.

In practice the GEV model is mis-specified, so the estimators have an
unknowable, albeit typically small, bias. Thus comparison purely in
terms of asymptotic variance is somewhat idealised, especially for the
return level, for which confidence intervals are typically highly
asymmetric. Notwithstanding this, our computations suggest that there
can be a substantial information loss when a sample of GEV variables is
replaced by its maximum, and thus strengthen the case for taking the
shortest blocks that lead to stable inferences. We now consider how to
assess this.

\section{Testing the block length}\label{testing-the-block-length}

\subsection{Basic idea}\label{basic-idea}

We suppose that the distribution of a block maximum is
\(\mathsf{GEV}(\mu_m', \sigma_m', \xi')\) and compare this to contributing maxima by performing a likelihood ratio test of the null
hypothesis that \(\mu_m'=\mu_m\), \(\sigma_m'=\sigma_m\) and
\(\xi'=\xi\). The success of this idea will depend on the alternative,
which we discuss after considering how the likelihood should be
constructed.

If \(X_{j} \iid \mathsf{GEV}(\mu, \sigma, \xi)\), we can use
max-stability to rewrite the marginal density of \(X_{(m)}\) in terms of
parameters \(\mu_m\) and \(\sigma_m\). The joint density of the \(m\)
observations is the product of the density of the block maximum and the
conditional density of the \(m-1\) smallest order statistics given the
maximum, which is \begin{align*}
f_{{-(m)} \mid {(m)}}(\boldsymbol{x}_{-(m)}) &= \frac{(m-1)!}{\sigma^{m-1}} \prod_{j=1}^{m-1} \frac{\exp \left\{-\left(1+\xi \frac{x_{(j)}-\mu}{\sigma}\right)_{+}^{-1/\xi}\right\}}{\exp \left\{-\left(1+\xi \frac{x_{(m)}-\mu}{\sigma}\right)_{+}^{-1/\xi}\right\}} \left(1+\xi \frac{x_{(j)}-\mu}{\sigma}\right)_{+}^{-1/\xi-1} \\
&= (m-1)!\prod_{j=1}^{m-1}\frac{f(x_{(j)}; \mu, \sigma, \xi)}{F(x_{(m)}; \mu, \sigma, \xi)}.
\end{align*} Up to permutation, the distribution of the smaller order
statistics given that \(X_{(m)}=x_{(m)}\) is that of \(m-1\) independent
GEV variables truncated above at \(X_{(m)}\).

For a random sample of \(n \times m\) observations, split into \(n\)
blocks of \(m\), we write the likelihood as \begin{align*}
\mathcal{L}({\boldsymbol{\vartheta}}_0, {\boldsymbol{\vartheta}}_m)=\prod_{i=1}^n f_{{-(m)} \mid {(m)}}(\boldsymbol{x}_{i, -(m)};{\boldsymbol{\vartheta}}_0)f_{(m)}(x_{i,(m)}; {\boldsymbol{\vartheta}}_m),
\end{align*} where \({\boldsymbol{\vartheta}}_0 = (\mu, \sigma, \xi)\)
and \({\boldsymbol{\vartheta}}_m\) are the parameters of the largest
observation within each block under the alternative, and where
\(\boldsymbol{x}_{i, -(m)} = (x_{i,(1)}, \ldots, x_{i,(m-1)})^\top\). If
the data are max-stable (or equivalently GEV distributed), then
\({\boldsymbol{\vartheta}}_m=(\mu_m, \sigma_m, \xi)\) is a bijective
transformation of \({\boldsymbol{\vartheta}}_0\). The idea underlying
our testing procedure is to let the parameters of the block maximum
differ, by
\begin{enumerate}
\def\labelenumi{\arabic{enumi}.}
\tightlist
\item
  fixing the shape and allowing the location and scale parameters to
  follow max-stability for block length \(\omega m\), with
  \(\mathcal{H}_a: {\boldsymbol{\vartheta}}_m = (\mu_{\omega m}, \sigma_{\omega m}, \xi)\)
  for some scaling \(\omega > 0\), and testing the null hypothesis that
  \(\omega=1\);
\item
  fixing the shape, but allowing both location and scale to vary with
  the block length, meaning
  \(\mathcal{H}_a: {\boldsymbol{\vartheta}}_m = (\mu_m+\nu, \phi \sigma_m, \xi)\)
  and testing the null hypothesis \((\nu=0, \phi=1)\); or
\item
  letting all three parameters vary, taking
  \({\boldsymbol{\vartheta}}_m = (\mu_m+\nu, \phi \sigma_m, \xi + \zeta)\)
  and testing the null hypothesis \((\nu=0, \phi=1, \zeta=0)\).
\end{enumerate}

These allow for varying degrees of flexibility, but since samples may be
small, the potentially mis-specified alternatives \(A_1\)--\(A_2\) using
few degrees of freedom may give more power. The form of alternative
\(A_1\) is suggested by the max-stability relationship under extremal
serial dependence; the shape parameter is hard to estimate, so taking it
to be the same for all the observations should provide some gain in
power. In practice we can expect the shape to vary only slowly, based on
the theory of penultimate approximations for distributions other than
the GEV, so fixing the shape as in alternative \(A_2\) seems a
benign compromise. Alternative \(A_3\) is the most flexible and is
useful when the shape varies quickly, but risks a loss of power.

\subsection{Censoring and rounding}\label{sec-censoring-rounding}

The generalized extreme value (GEV) distribution is continuous, and
might provide a poor approximation for maxima in some cases. For
example, air pollution or precipitation maxima can sometimes be very
small, even when taken over sizable periods, and this can strongly
influence the fitted distribution. In a likelihood-based framework and
when the focus is on the upper tail of the distribution, such issues can
be accommodated by left-censoring any measurement below some level
\(u\). Moreover, historical data may be rounded so
coarsely that it is unwise to treat them as continuous. In this section
we describe how the likelihoods may be adjusted accordingly.

Left-censoring at \(u\) is handled simply by replacing the density
function with the cumulative distribution function evaluated at \(u\). We deal in addition with rounding to the nearest \(\delta\) by
considering the likelihood of the latent variables
\(X^*_{i,j} \sim \mathsf{GEV}(\mu, \sigma, \xi)\), where we observe
rounded measurements \(X_{i,j}=x_{i,j}\), where
\(X_{i,j} - \delta/2 \le X^*_i \le X_{i,j}+ \delta/2\); the
contributions of individual observations to the likelihood when
\(x_{i,(j)} > u\) are
\begin{align*}
L_{i,j}(\boldsymbol{\vartheta}) = \begin{cases}
f(x_{i,(j)}; \boldsymbol{\vartheta}), & \delta=0, \\
F(x_{i,(j)}+\delta/2; \boldsymbol{\vartheta})- F(\max\{x_{i,(j)}-\delta/2, u\}; \boldsymbol{\vartheta}), & \delta > 0.
\end{cases}
\end{align*} The joint likelihood contributions for the \((m-1)\) lowest
order statistics of \(m\) are from the GEV distribution right-truncated
at \(x_{(m)} + \delta/2\).

When rounding and left-censoring are combined, the likelihood
contribution from \(x_{i,(1)}, \ldots, x_{i,(m)}\) is \begin{multline*}
L_i(\boldsymbol{\vartheta}_0, \boldsymbol{\vartheta}_m) = L_{i,m}(\boldsymbol{\vartheta}_m)^{w_{i,(m)}(\boldsymbol{\vartheta}_m)} F(u; \boldsymbol{\vartheta}_m)^{1-w_{i,(m)}(\boldsymbol{\vartheta}_m)} \\ \times \prod_{j=1}^{m-1} \frac{L_{i,j}(\boldsymbol{\vartheta}_0)^{w_{i,(j)}(\boldsymbol{\vartheta}_0)} F(u; \boldsymbol{\vartheta}_0)^{1-w_{i,(j)}(\boldsymbol{\vartheta}_0)}}{F \left(\max\{u, x_{i,(m)} + \delta/2\}; \boldsymbol{\vartheta}_0\right)}
\end{multline*} where the weight is
\begin{align*}
w_{i,j}(\boldsymbol{\vartheta}) = \mathsf{P}_{\boldsymbol{\vartheta}}(\max\{u, x_{i,(j)}- \delta/2\} \leq  Y_{i,(j)} \le x_{i,(j)} + \delta/2)/ \mathsf{P}_{\boldsymbol{\vartheta}}(|Y_{i,(j)} -x_{i,j}|\le\delta/2).
\end{align*}
This is equivalent to use of an expected likelihood
\citep[Section 2.2 of][]{Varty:2021}. The overall likelihood is
\(\prod_{i=1}^n L_i(\boldsymbol{\vartheta}_0, \boldsymbol{\vartheta}_m)\).

\section{Simulation study: power
analysis}\label{simulation-study-power-analysis}

\subsection{General setup}\label{general-setup}

Below we use simulation to study the properties of our proposed tests
for departures from max-stability, using sample sizes
\(n \in \{25,50,100\}\) replicates of \(m\in \{2, 5, 10\}\) observations
and three different alternatives. We obtain \(p\)-values by comparing the likelihood ratio statistic with its asymptotic null
distribution, which is chi-square with one, two and three degrees of
freedom under the three alternatives. All power curves displayed in this
section are based on 2000 replications, with tests conducted at level
5\%.

The penultimate approximation to the GEV distribution is already very good
for samples of size 30: the GEV densities from the penultimate
approximations to the Weibull or normal are nearly indistinguishable
from those from \(G^{30}\). We therefore defer to the
supplementary material results for so-called max-domain of attraction simulations, i.e., those from
from \(G^{30}\). These show similar qualitative behaviour to those
presented in Section~\ref{sec-gev-power}.

\subsection{Simulation from generalized extreme value
models}\label{sec-gev-power}

The first scenario is an alternative where the sample of \((m-1)\)
lowest order statistics are
\(\mathsf{GEV}(\mu_0 = 0, \sigma_0 = 1, \xi_0 = 0.1)\), but the largest
observation is generated from a GEV distribution conditioned to lie
above the \((m-1)\)th order statistic and with parameters
\(\mu = \mu_0 + \delta\), \(\sigma=\sigma_0 \exp(-\delta/10)\) and
\(\xi = \xi_0\). The rationale for this alternative is that the
maximum likelihood estimators of the GEV location and scale parameters
have an asymptotic positive correlation of \(0.48\) based on the inverse
Fisher information for the parameter values. Thus, alternatives that
show negatively correlated changes to \((\mu_1, \sigma_1)\) should be in
principle easy to detect.

Scenarios two and three compare two GEV distributions whose parameters
are derived by considering penultimate parameters obtained from the
maximum of \(30 \delta\) independent and identically distributed
standard normal draws, and those from a Weibull distribution with unit
scale and shape \(0.8\), with \(\delta \ge 1\). For each parameter
combination, we generate \(n\) samples of \(m\) largest order statistics
from the approximating GEV distribution, and use the conditional
simulation described for the first scenario to ensure that the largest
of the \(m\) observations exceeds the others. All of the parameters are
strongly correlated, but the shape parameter variability leads to more
detectable changes as \(m\) increases, especially for low and high
quantiles of the underlying sample.

\begin{figure}[tbp!]

\centering{

\includegraphics[width=\textwidth]{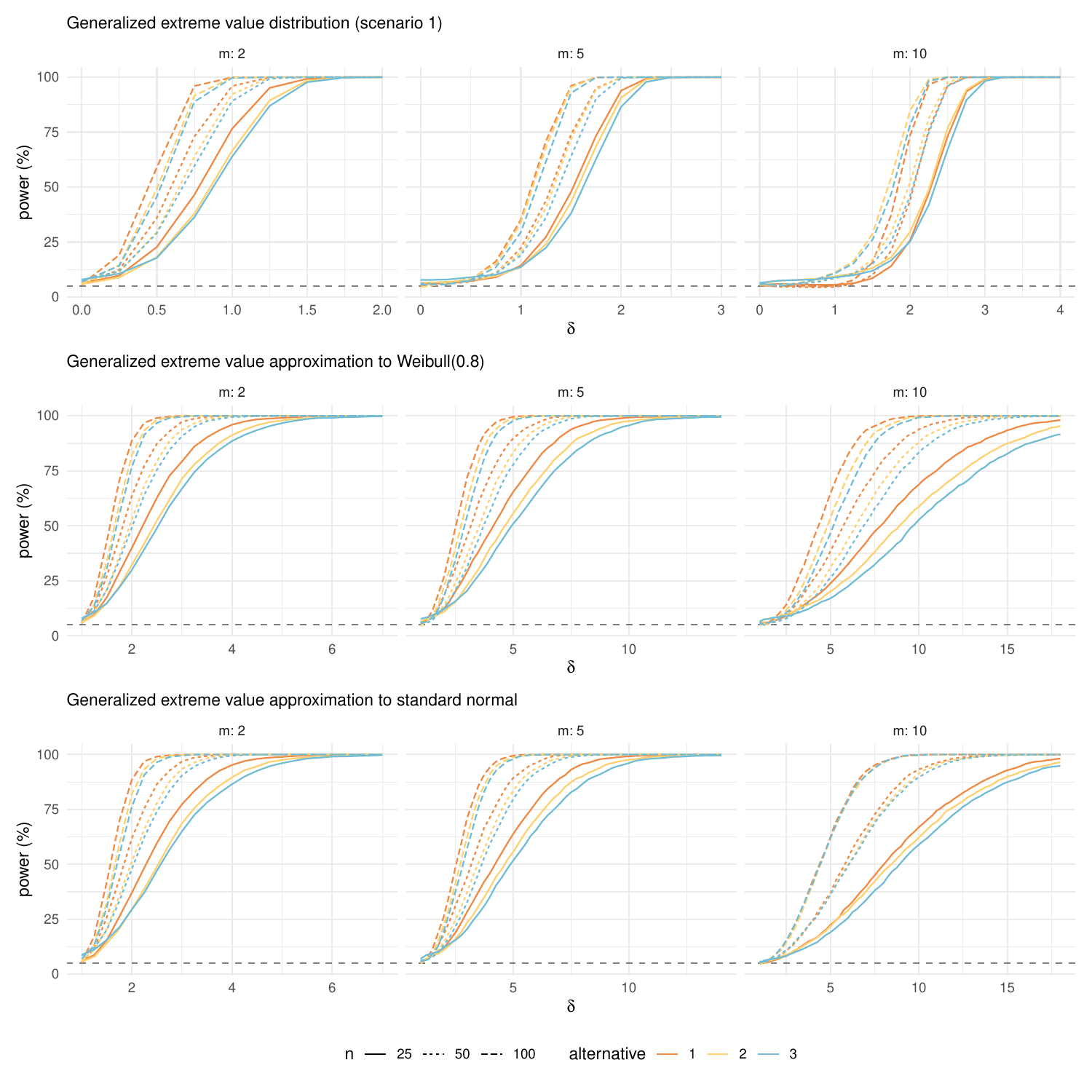}

}

\caption{\label{fig-power-curves}Power curves for simulations from the
GEV model (top), a GEV whose parameters are based on a Weibull with
shape 0.8 (middle) and a standard normal penultimate model (bottom),
both of the latter from maxima of 30 draws. The power is obtained from
repeated samples of size \(n \times m\). Under the alternative, we vary
\(\delta\), with values of \(0\) (top) and \(1\) (bottom two rows)
corresponding to the null hypothesis of max-stability.}

\end{figure}

Figure~\ref{fig-power-curves} shows the power curves for the three GEV distributions considered. In all cases, the most powerful option is to use
alternative \(A_1\), even when that alternative hypothesis is
mis-specified, as in the first setting. Power is higher when the sample
size grows and when the alternative departs further from the null model.
Perhaps surprisingly, power is also higher for smaller values of \(m\), maybe because differences between the two largest order statistics can be larger when $m$ is small.

The size is distorted by up to around 1.5\% for \(m=5\) for alternatives
with two or three parameters, at sample sizes \(n \in \{25, 50\}\). This
can be further assessed by testing departures from uniformity based on
Kolomogorov--Smirnov tests. The size distortion disappears as either
\(m\) or \(n\) increase.

\subsection{Simulations from stationary
sequences}\label{simulations-from-stationary-sequences}

To investigate the effect of serial dependence, we simulate stationary
time series based on first-order max-autoregressive processes
parametrized in terms of the extremal index \(\theta\), whose reciprocal
can be interpreted as the average cluster length \citep[cf.~Section
10.2 of][]{Beirlant:2004}.

For \(\xi=0\), we follow \citet{Tavares:1977}: let
\(\theta \in (0,1]\) and consider data obtained through the recursion
\(Y_i = \max\{Y_{i-1}, Z_i\} +\log(1 - \theta)\) for \(i \geq 1\), where
the innovations
\(Z_i \sim \mathsf{GEV}\{\log(\theta) - \log(1-\theta),1,0\}\) and the initial state
\(Y_0 \sim \mathsf{GEV}(0,1,0)\) is standard Gumbel. It is then easily
shown that the marginal distribution of \(Y_i\) is standard Gumbel. The
maximum of \(m\) observations from the stationary sequence is
Gumbel with location \(\log[1+(m-1)\theta]\).

For positive shape parameter, \(\xi>0\), we instead simulate recursively
from the first-order max-autoregressive (MAR) model
\begin{align*}
Y_i= \max\{(1-\theta)^\xi Y_{i-1}, Z_i\}, \qquad i =1, 2,\ldots;
\end{align*} the innovations
\(Z_i \sim \mathsf{GEV}(\theta^{\xi}, \xi\theta^{\xi}, \xi)\) are
independent and identically distributed and \(Y_0\) is drawn from the
marginal stationary distribution of the process,
\(F = \exp(-X^{-1 / \xi})\), i.e., \(Y_0\) has a
\(\mathsf{GEV}(1, \xi, \xi)\) distribution. This yields a stationary
sequence with extremal index \(\theta\). The distribution of
\(\max\{Y_i, \ldots, Y_{i+m}\}\) from this process is Fréchet with
distribution function \(\exp[-\{\theta(m-1)+1\}x^{-1/\xi}]\); see
Example 10.3 of \citet{Beirlant:2004}.

\begin{figure}[tbp!]

\centering{

\includegraphics[width=\textwidth]{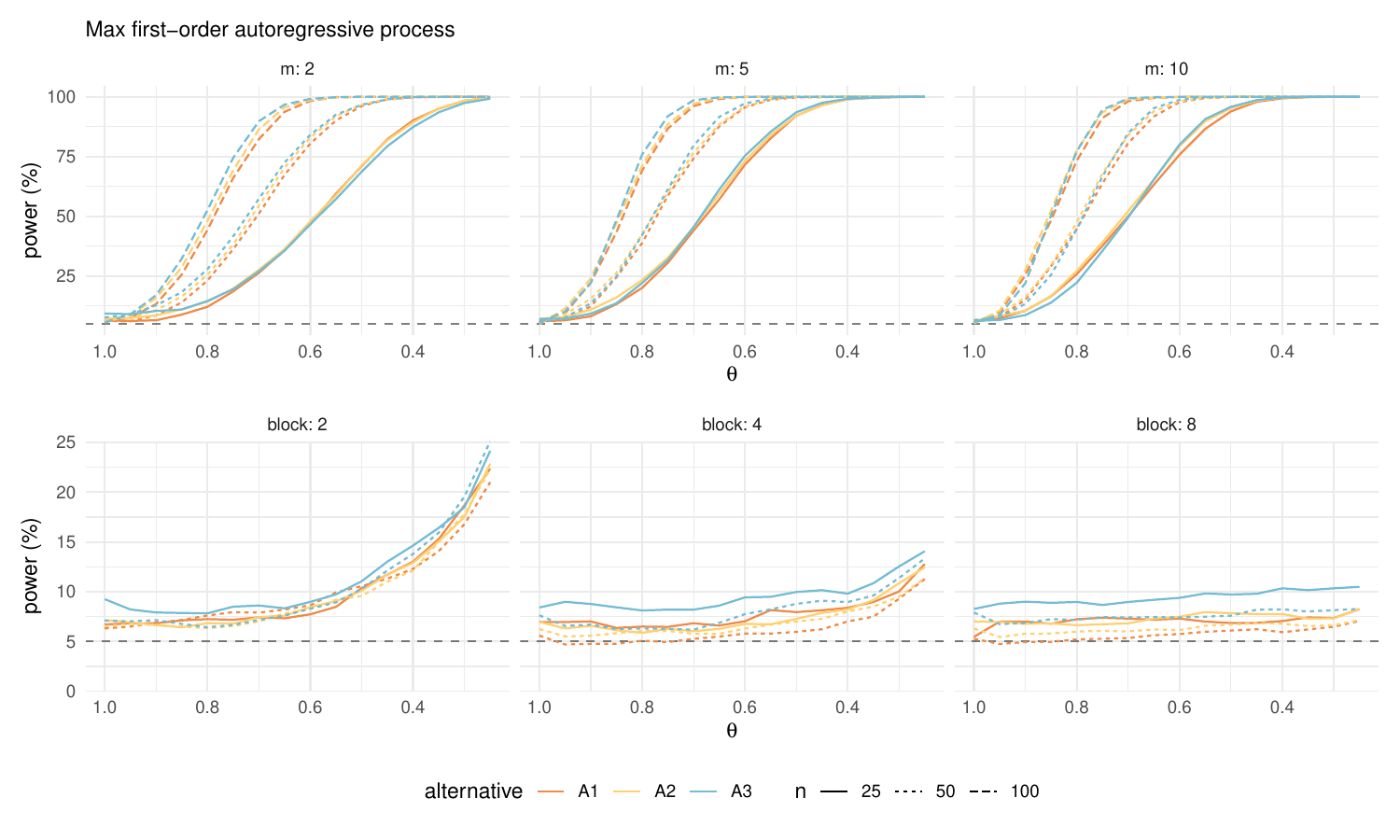}

}

\caption{\label{fig-maxautoregressive}Power curves for simulations from
first-order max-autoregressive processes. The top panel show the effects
of changing the number of observations evaluated (\(m \in \{2,5,10\}\),
left to right). The bottom panel shows the impact of fitting GEV to
maxima of blocks of length \(m\in\{2,4,8\}\) (left to right) from the
max-autoregressive processes, comparing with a doubling of the sample
size \(m_2=2m_1.\) Results are averages over shape parameters
\(\xi \in\{0, 0.2, 0.4\}\), and displayed as a function of the extremal
index \(\theta\) (reversed \(x\)-axis).}

\end{figure}%

In the case of max-autoregressive processes, the data have a marginal
GEV distribution, but due to serial dependence the likelihood is
mis-specified; max-stability only holds approximately, and with $m$ replaced by  \(m\theta\).  We took \(\xi \in \{0, 0.2, 0.4\},\),  but a binomial analysis of deviance of the results showed no effect of the shape parameter, so we marginalize over it and present pooled
estimates. The top line of Figure~\ref{fig-maxautoregressive} shows
power curves as the dependence increases, i.e., the
extremal index \(\theta \to 0\) towards the right (note the reversed
\(x\)-axis). The power rises as \(m\) increases: all three
tests have similar power, as every alternative is true but only the
parameters vary. We again see slight size distortion in small samples
for the more flexible alternative \(A_3\), which explains its ostensibly higher power.

When dealing with stationary time series in practice, one would
effectively increase the block length to get more or less independent
block maxima if the serial correlation is strong. As the block length
increases, the GEV approximation should become appropriate and will
directly incorporate the impact of extremal clustering. However, since
the range of the extremal dependence is necessarily finite, comparing
the model fitted to blocks of length \(m_1\) and \(m_2\), with
\(m_1< m_2\) still should lead to small discrepancies, because maxima of the longer blocks would be closer to independence, and thus series
of maxima of blocks of length \(m_2\) would lead to lower estimates of
\(\theta\). The bottom line of Figure~\ref{fig-maxautoregressive} shows
that, as we would expect,  the test would not detect departures from the null for our
first-order max autoregressive process for longer blocks and when the average cluster length \(\theta^{-1}\) is not too large.
As the block length increases, the effect of serial dependence vanishes
as expected. Alternative \(A_3\) suffers again from size distortion,
especially in small samples.

\section{Graphical diagnostics}\label{graphical-diagnostics}

\subsection{General}\label{general}

Diagnostic plots are valuable counterparts to the quantitative
information provided by the result of a statistical test. In the present
case there are numerous ways of measuring departure from the model, but
in light of our alternatives, we focus on the block maximum \(X_{(m)}\)
and the set of all block maxima, suitably transformed to be approximate
uniform pivots. Quantile-quantile plots are widely used to assess
goodness-of-fit, but their interpretation is greatly bolstered by
indications of uncertainty; we propose the use of a parametric bootstrap
scheme, if necessary making allowance for rounding and truncation of the
observations \citep[cf.][]{Varty:2021}. To obtain pointwise and
simultaneous confidence intervals using the bootstrap, one option is the
envelope method \citep[section 4.2.4 of][]{Davison.Hinkley:1997}, but
we instead borrow ideas from \citet{Sailynoja.Burkner.Vehtari:2022}
and use binomial confidence intervals.

For the bootstrap we consider, we seek pivotal quantities
whose null distribution does not depend on parameters. The simplest of
these are based on the uniform distribution of the independent variables
\(F(X_{j}; \boldsymbol{\vartheta})\), or on
\(F(X_{(m)}; \boldsymbol{\vartheta}_m)\), or equivalently on
\(F(X_{(m)}; \boldsymbol{\vartheta}) \sim \mathsf{beta}(m,1)\).

\subsection{Parametric bootstrap for confidence
intervals}\label{sec-parametric-bootstrap}

Section 2.2 of \citet{Sailynoja.Burkner.Vehtari:2022} outlines a
method for building simultaneous confidence intervals for tests of
uniformity in probability-probability plots, evaluated at \(N\) fixed
plotting positions \(\nu_1, \ldots, \nu_N \in (0,1)\). They consider the
values of the empirical distribution function (ECDF)
\(\mathsf{ECDF}(\nu; \boldsymbol{v}) = n^{-1} \sum_{i=1}^n \mathsf{1}_{\nu \leq v_i}\)
for \(\boldsymbol{v} \in [0,1]^n\) and \(\nu \in (0,1)\). It is easy to
see that, with a uniform sample, \(n\mathsf{ECDF}(\nu; \boldsymbol{v})\)
is binomial with \(n\) trials and probability of success \(\nu\), whose
distribution function we denote \(H(\cdot; n, \nu)\): we can obtain a
pointwise confidence interval at any value \(\nu\) from the quantile
function of the binomial and scale the latter by \(n\).

To obtain simultaneous confidence intervals,
\citet{Sailynoja.Burkner.Vehtari:2022} consider simulation of \(B\)
uniform samples of size \(n\), say
\(\boldsymbol{v}_b \in [0,1]^n (b=1, \ldots, B)\) and finding for the
\(b\)th bootstrap the minimum level \(\alpha_b\) such that the entire
scaled ECDF lies inside the interval, where
\begin{equation}\protect\phantomsection\label{eq-alpha-boot}{
\alpha_b = 2 \min_{i=1}^n \left\{ \min\left\{H\left[n\mathsf{ECDF}(\nu_i; \boldsymbol{v}_b); n, \nu_i\right], 1 - H\left[n\mathsf{ECDF}(\nu_i; \boldsymbol{v}_b)-1; n, \nu_i\right]\right\}\right\}.
}\end{equation} To obtain simultaneous intervals with level of
significance \(\alpha\), they instead draw intervals with
\(\alpha^*\) the \(\alpha\) quantile of
\(\alpha_1, \ldots, \alpha_B\). The pointwise and simultaneous
confidence intervals can be calculated for the original sample at
\(\boldsymbol{\nu}\) through
\([H^{-1}(\alpha/2; n, \nu_i), H^{-1}(1-\alpha/2; n, \nu_i)]\) and
\([H^{-1}(\alpha^*/2; n, \nu_i), H^{-1}(1-\alpha^*/2; n, \nu_i)]\).

Their method, unadjusted,  fails when we instead use approximate uniform draws from pivots
\(F(\boldsymbol{X}; \widehat{\boldsymbol{\vartheta}})\). To illustrate this, we generated 1000 samples of 100 independent
uniform samples and calculated 50\% simultaneous confidence interval
using the simulation-based method: their coverage was 50.5\%. We
repeated the procedure with standard Gumbel variates (\(\xi=0\)), and
transformed these using the probability integral transform with maximum
likelihood estimates in place of the parameters, giving coverage
97.6\%. We can fix this by taking into account the estimation
uncertainty and potentially different sample sizes through the bootstrap
by replicating the following procedure, whose adequacy is discussed in the supplementary material.

A conventional parametric bootstrap for a sample of size \(n\), from
which the maximum likelihood estimate
\(\widehat{\boldsymbol{\vartheta}}\) is obtained,
proceeds as follows:

\begin{enumerate}
\def\labelenumi{\arabic{enumi}.}
\tightlist
\item
  transform $nm$ independent uniform draws
  to GEV variables by applying the quantile
  function \(F^{-1}(\cdot; \widehat{\boldsymbol{\vartheta}})\), and
  order them in an $n \times m$ matrix to obtain order statistics
  \(X^{*}_{i,(1)},\ldots X_{i,(m)}^{*}\) in row $i=1, \ldots, n$;
\item
  calculate an estimate \(\widehat{\boldsymbol{\vartheta}}^{*}\) from
  the ordered sample using the same estimation method;
\item
  map the bootstrap sample to the uniform scale using the probability
  integral transform
  \(F(X_{j}^*;\widehat{\boldsymbol{\vartheta}}^{*})\) or \(F(X_{j}^*;\widehat{\boldsymbol{\vartheta}}_m^{*})\) for the row maximum;
\item
  compute the empirical distribution function of the resulting uniform
  sample, evaluate it at a set of \(N\) predetermined uniform
  plotting positions \(\nu_1=1/N, \ldots, \nu_{N-1} = (N-1)/N\), and
  obtain \(\alpha_b\) from eq.~\eqref{eq-alpha-boot}.
\end{enumerate}

The procedure is repeated \(B\) times and we obtain the overall level
\(\alpha^* < \alpha\) for simultaneous confidence intervals.

A difficulty with this bootstrap procedure is that the null distribution
of any statistic based on the bootstrap data will depend on the
parameter value \(\widehat{\boldsymbol{\vartheta}}\) used for simulation, which differs from the true value
\(\boldsymbol{\vartheta}\), and although
\(\widehat{\boldsymbol{\vartheta}}\) is consistent under the null
hypothesis, some distortion of the null distribution will remain. The
resulting error could be reduced by nested bootstrapping \citep[section
3.9]{Davison.Hinkley:1997}, but as the first-order bias of the maximum
likelihood estimator for the GEV distribution is negligible and
simulation studies (not reported) indicated little to no impact of the
bias when using implicit bootstrap bias correction, we do not pursue
this further.

\subsection{Bootstrap with censoring and
rounding}\label{sec-bootstrap-censoring-rounding}

The bootstrap schemes and quantile-quantile plots described above must
be modified when the data are rounded or left-censored below \(u\), as
shown by \citet{Varty:2021}. Here we outline how these issues are handled, noting that
estimation of parameters for the bootstrap sample with rounded and
censored data under the null hypothesis is straightforward.

The exact values of rounded observations are unknown and, although they
could be represented as vertical segments in a quantile-quantile plot,
rounding often leads to ties and poses further problems for the
construction of pivots, which are generally only approximate when based
on discrete observations. The latent observations can be imputed from
their conditional distribution, by sampling
\(Y_{i,j} \mid X_{i,j}= x_{i,j}\) from a GEV distribution restricted to
the interval \([x_{i,j} - \delta/2, x_{i,j} + \delta/2]\). We can then
calculate pivots as usual. This form of randomized data augmentation has
the drawback that different draws lead to different plotting positions.

A second problem is due to left-censoring: under the bootstrap scheme,
\(n\) observations are generated on the data scale, but they are
left-censored if they fall below the lower threshold \(u\), leading to a
random number of exceedances, \(n^{*}_u\), say, above \(u\).
Left-censored points do not appear in the quantile-quantile plot, so the
plotting positions must be adjusted by treating the GEV distribution as
left-truncated at \(u\). \citet{Varty:2021} resolves this problem of
unequal sizes by transforming the \(n^{*}_u\) observations above \(u\)
to a standardized scale, calculating their empirical quantile function,
and then evaluating the latter at \(N\) plotting positions: this ensures
that the uncertainty is captured. This is similar in spirit to
\citet{Sailynoja.Burkner.Vehtari:2022} choosing a grid of uniform
plotting positions.

For left-censored or interval-censored data, the steps of the parametric
bootstrap are therefore modified as follows: in step 1, the measurements
are rounded to the same precision as the original data and treated as
interval-censored, and left-censored if they fall below \(u\) in the
likelihood in step 2. Step 3 is only applied to the continuous data
imputed from the corresponding truncated GEV on
\([x_{j}^* -\delta/2, x_j^* + \delta/2]\), reordered within each block
of length \(m\). Any point simulated below \(u\) is left-censored and
discarded from the plotting positions, leaving \(n^{*}_{u} < n\)
distinct values for the quantile-quantile plot. Finally, a final step
takes the bootstrap sample of uniform draws obtained from the pivots, of
size \(n_u^{*}\), calculates the empirical distribution function of
these and evaluates them at the set of predetermined \(N\) uniform
plotting positions.

\section{Applications}\label{applications}

\subsection{Hourly maximum wind speed at
Cheeseboro}\label{application-1-hourly-maximum-wind-speed-at-cheeseboro}

We revisit the data presented in Section~\ref{sec-motivating-example}, which were obtained from the station daily summary. Hourly maxima of these data were analyzed by
\citet{Reich.Shaby:2016}, but there is clear evidence of non-stationarity at the hourly (versus
daily) level, so we focus on daily values. Some hourly measurements are missing but there are data for every day. We use a
discrete likelihood assuming rounding to the nearest integer, though
this only changes the third decimal place of maximum likelihood
estimates for \(m=4\), as small rounding has limited on
point estimates (but more on their uncertainty).

\begin{figure}[tbp!]

\centering{

\includegraphics[width=\textwidth]{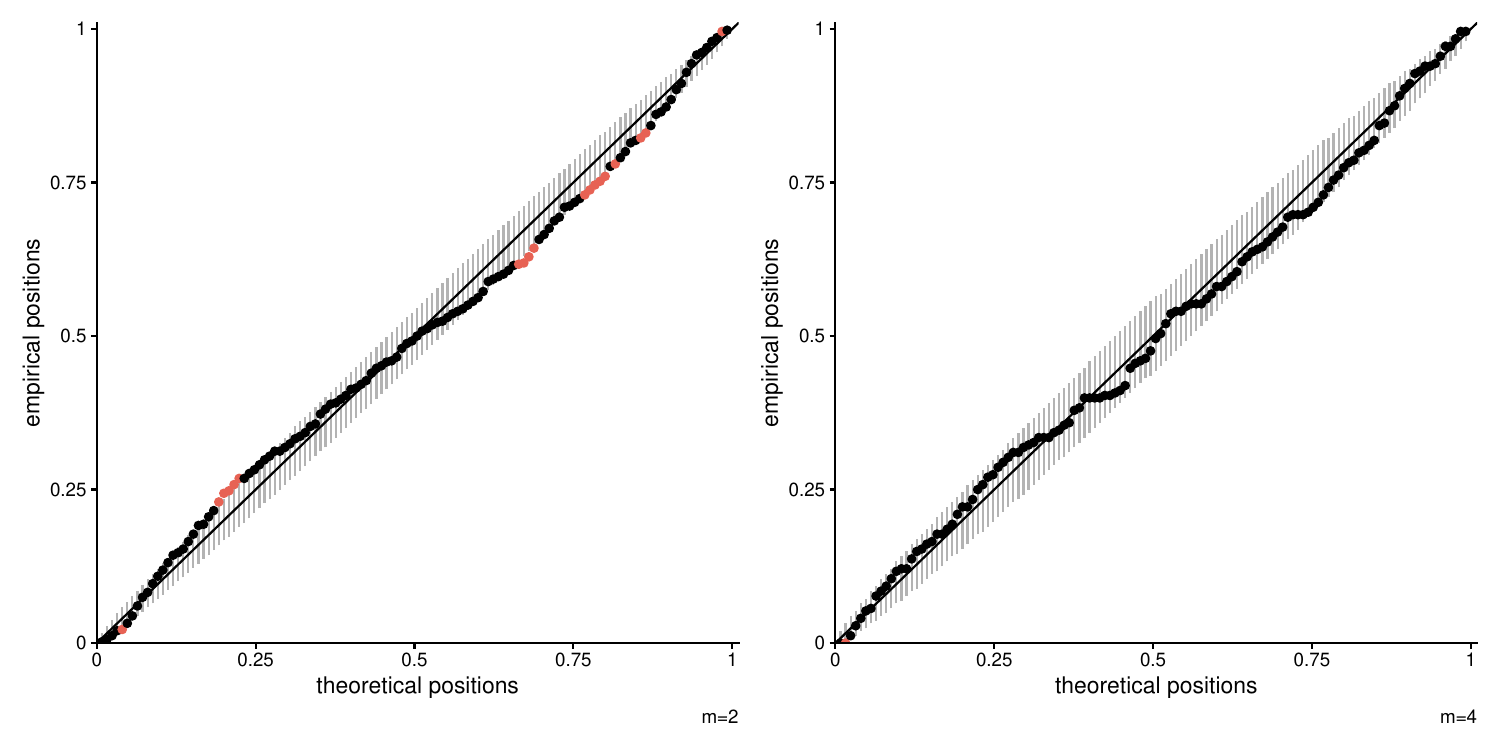}

}

\caption{\label{fig-qqplots-cheeseboro}
Probability-probability plots for block maxima based on two-day
(left) and four-day (right) block lengths, with 95\% simultaneous
binomial-based confidence intervals. Points drawn in red fall outside of the simultaneous intervals.}

\end{figure}

\begin{table}[tbp!]

\caption{\label{tbl-pvals-cheeseboro}P-values (rounded to three digits)
for max-stability test for blocks of size \(m\) days versus \(cm\)
days.} \label{tbl-test-Cheeseboro}

\centering{

\centering
\begin{tblr}[         
]                     
{                     
colspec={Q[]Q[]Q[]Q[]Q[]Q[]Q[]},
hline{2}={3}{solid, black, 0.03em},
hline{2}={2,5}{solid, black, 0.03em, l=-0.5},
hline{2}={4,6}{solid, black, 0.03em, r=-0.5},
hline{3}={1-7}{solid, black, 0.05em},
hline{1}={1-7}{solid, black, 0.08em},
hline{7}={1-7}{solid, black, 0.08em},
cell{1}{1}={}{halign=c},
cell{1}{2}={c=3}{halign=c},
cell{1}{3}={}{halign=c},
cell{1}{4}={}{halign=c},
cell{1}{5}={c=2}{halign=c},
cell{1}{6}={}{halign=c},
cell{1}{7}={}{halign=c},
}                     
& \(c=2\) &  &  & \(c=4\) &  &  \\
block & \(A_1\) & \(A_2\) & \(A_3\) & \(A_1\) & \(A_2\) & \(A_3\) \\
\(m=1\) & 0 & 0 & 0 & 0 & 0 & 0 \\
\(m=2\) & 0.014 & 0.047 & 0 & 0.13 & 0.3 & 0 \\
\(m=4\) & 0.383 & 0.546 & 0.16 & 0.2 & 0.41 & 0.18 \\
\(m=8\) & 0.137 & 0.318 & 0.43 & 0.66 & 0.76 & 0.81 \\
\end{tblr}

}

\end{table}%
$P$-values for the test of max-stability are reported in Table~\ref{tbl-test-Cheeseboro}: comparing daily (versus two
days) and two days versus four days, we find strong evidence against
max-stability when using either
\(c=2\) or \(c=4\). While this doubling procedure might seem an
artificial way to choose the block length, the quantile-quantile plots
in Figure~\ref{fig-qqplots-cheeseboro} show that the extrapolation of
the fit from daily or two-day maxima do not capture longer time
horizons, whereas the estimates from blocks of length
\(m\in \{4,8,16\}\) appear more or less equivalent.
The GEV approximation is poor when \(m=1\) (overestimating the 20
largest observations by a wide margin) and, as shown in the left-panel of Figure~\ref{fig-qqplots-cheeseboro}, not quite appropriate for
\(m=2\).
This is compatible with the behaviour observed in Figure~\ref{fig-density-50year}.

If we fit the GEV model to blocks of size \(m\) and get maximum
likelihood estimates
\((\widehat{\mu}, \widehat{\sigma}, \widehat{\xi})\), and there are
\(n_m\) blocks per year, then max-stability implies that the maximum
likelihood estimator of the distribution of the 50-year maximum, and any
resulting risk functional such as its mean or median, is obtained from
\(F(\widehat{\mu}_{n_mN}, \widehat{\sigma}_{n_mN}, \widehat{\xi})\).
The maximum likelihood estimators for the GEV model fitted to blocks of
length \(m=4\) days can be employed to calculate risk measures for the
50-year maximum, by using the max-stability relationship with
\(m=8 \times 50\) and calculating quantiles or moments from the
resulting distribution. Confidence intervals can be obtained by
profiling the model in terms of the resulting functional, yielding point
estimate for the median 50-year maximum (95\% confidence interval) of
\(86.73 (83.06,105.07)\) miles per hour.

This calculation does not account for the serial clustering: the
extremogram (not shown) reveals no dependence at any lag, but the series
is short and the estimated extremal dependence depends on the threshold
(for threshold-based estimators) or the block length, though these do
not account for the gaps in the time series across years. The maximum
likelihood estimator of \(\theta\) based on \citet{Suveges:2007} with
a marginal threshold at the 0.95 quantile is 0.82. Adjusting for this
gives an estimate of \(85.45\) miles per hour, close to the previous
estimate.

\subsection{Flow of the River Thames}\label{application-2-river-flow-of-the-thames-at-kingston}

The flow of the Thames at Kingston has been recorded since
1883. The data, which were obtained from the
\href{https://nrfa.ceh.ac.uk/data/station/info/39001}{National River
Flow Archive} (site 39001), contain 142 annual measurements, of which 32
are marked as natural annual maximum mean daily flows rather than
instantaneous annual maxima. The value for the year 1894 was backcast,
as the previous value was considered to overestimate the true value.

We can compare the fits of the GEV for annual and biyearly maximum: the
coefficients for the two-year maximum implied by max-stability based on
fitting the GEV to annual maxima are \(\mu_2=341.52\),
\(\sigma_2=92.39\) and \(\xi_2=-0.06\), whereas those obtained from the
smaller data of biyearly maximum are \(\widehat{\mu}_2=338.26\),
\(\widehat{\sigma}_2=96.6\) and \(\widehat{\xi}_2=-0.06:\) the shape
appears roughly to be constant. Estimates of the extremal index equal
unity, indicating no clustering of extremes for the annual maxima. The
\(p\)-values for the likelihood
ratio tests with annual maxima compared with biyearly \((m=2)\) are
\(0.44\), \(0.63\), and \(0.14\), for alternatives \(A_1\), \(A_2\) and
\(A_3\), respectively. None suggests any departure from
max-stability, so using annual maxima seems adequate.

\subsection{Abisko rainfall and
landslide risk}\label{application-3-abisko-rainfall-and-landslide}

The Abisko scientific research station, located in Lapland (Sweden), has
experimental and observational records from January
1913 until December 2014. The surrounding region is prone to
landslides due to prolonged heavy rainfall.
\citet{Kiriliouk:2019} analyse one-, two- and
three-day episodes using a multivariate generalized Pareto distribution
to assess landslide risk. \citet{Guzzetti:2007} identify
intensity-duration combinations  likely to lead to landslides,
and use raw
intensity-duration data for highland terrain to propose a nonlinear model \(r=7.56 \times h^{0.56}\) that relates the landslide-triggering threshold
\(r\) to hourly rainfall duration \(h\). For three-day durations, \((h=72)\), this threshold corresponds to
cumulated rainfall exceeding 69.9mm.

We compute three-day running sums and consider rainfall precipitation
with 12 periods of seven days from June 15th every year (approximately
three months) to ensure approximate stationarity. One problematic aspect
of rainfall  is block maxima near zero as a result of dry days,
causing strongly positive estimates of the shape parameter. Since we focus on extreme events, we left-censor such records below  10mm for three day episodes. The proportion of points censored is 0.62 for weekly
maxima, 0.19 for monthly and none for yearly. We consider as quantity
of interest the probability of an annual record exceeding 69.9mm.

\begin{figure}[tbp!]

\centering{
\includegraphics[width=\textwidth]{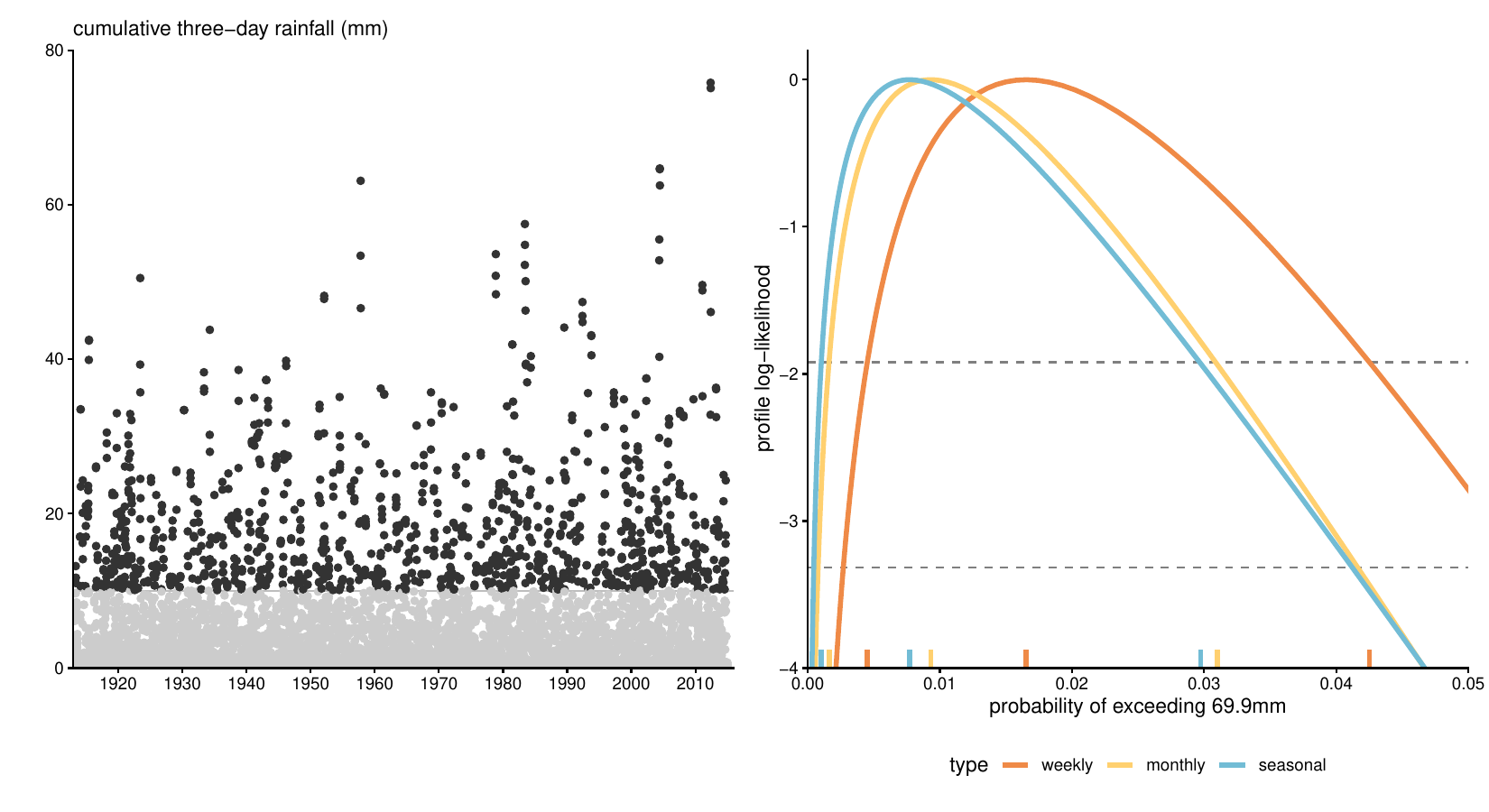}

}

\caption{\label{fig-plots-Abisko}Time series of three-day cumulated
rainfall measurements (in mm), with points above the threshold at
\(u=10\)mm higlighted in black (left), and profile likelihood for the
probability of having a summer rainfall exceeding 69.9mm based on the
GEV model fitted to weekly, 28-days and seasonal maxima (right). The
dashed horizontal lines give cutoff values for the 95\% and 99\%
asymptotic confidence interval limits. The rugs at the bottom indicate
the limits of the 95\% confidence interval and maximum likelihood
estimates.}

\end{figure}

P-values for the test of max-stability comparing 7-day to 28-day maxima
(\(m=4\)) yields \(p\)-values less than \(10^{-3}\).  The maximum
likelihood estimate of the shape parameter reduces by \(0.06\) for the
longer period. Comparing 28-day maximum to seasonal (\(m=3\))
gives \(p\)-values of \(0.69\) for alternative \(A_1\), \(0.69\) for
\(A_2\) and \(0.95\) for \(A_3\). Thus, monthly blocks appear
sufficient. The probability-probability plots (not shown) indicate
that both weekly and monthly data do not reject the null of adequacy at level 5\%,
provided small values are left-censored. The left panel of
Figure~\ref{fig-plots-Abisko} shows two empirical
exceedances, but they are serially correlated due to the running sum:
only the longer block seemingly captures the extremal clustering,
explaining the difference in point estimates for the target exceedance probability.
A sensitivity analysis varies $u$ to check the impact of the
left-censoring bound: the $p$-values decrease as $u$ increases,
but lead to the same conclusion and are of similar order of magnitude.
Higher values of $u$ lead to more negative shape estimates.
%
%
%
%
%

The fitted GEV distribution can be used to estimate the probability
that the seasonal maximum exceeds 69.9mm. The maximum likelihood estimate (95\%
profile-based confidence interval) based on \(m=28\) days is
\(9.3 \times 10^{-3}\) \([1.6 \times 10^{-3}\), \(31 \times 10^{-3}]\);
those for the fit based on the yearly or summer maximum based on blocks
of size \(m=28 \times 3\), with probability \(7.7 \times 10^{-3}\)
\([1 \times 10^{-3}\), \(29.7 \times 10^{-3}]\).
Figure~\ref{fig-plots-Abisko} shows that the maximum likelihood estimate
of the probability of exceedance is rather larger, again highlighting
how differences appear for the risk summaries of interest
rather than at the data level.

\section{Discussion}\label{discussion}

This paper proposes likelihood-based tests for max-stability to
determine the block length \(m\) when using the block maximum method. Such procedures were unnecessary when data were
generally available only at low frequencies, but have become more
relevant as higher-frequency data become more readily available and
maxima might be computed with many different block lengths. We compare
maxima with the distributions implied by smaller blocks, and use three
alternative distributions that are inspired by max-stability and penultimate approximations. Our best approach overall uses a one-dimensional alternative and has high power
even in small samples (with \(m=2\) and \(n=25\)), suggesting that it
can pick up discrepancies when extrapolating the distribution of
maxima. In the application to the Cheeseboro wind speed data, all three
alternatives suggested that blocks of lengths less than four
were unsuitable for risk assessment. The choice of \(m\) should
generally dictated by practical considerations linked to
interpretability, supported by procedures such as ours.

An obvious limitation is that our approach is restricted to independent
or stationary time series. This may require picking a suitable time
window, as we did for Cheeseboro wind speed, or pre-processing the
series and modelling the resulting residuals. Situations in which the
GEV parameters vary with covariates \citep[e.g.,][]{Davison.Smith:1990,Youngman:2022}
would complicate testing, not least because it is unclear whether
max-stability in such cases should be valid marginally or conditionally.

For stationary data, it is possible to estimate the extremal index
\(\theta\) directly and perform the test with \(m\) replaced by
\(m\theta\), but this invalidates the asymptotic theory we use. Provided the blocks are long enough for their
maxima to be considered independent, separate estimation of the extremal
index \(\theta\) is unnecessary when dealing with summaries from the
maximum of longer blocks, since the maximum likelihood estimator
captures the serial dependence through scaling parameters
\citep{Bucher.Zhou:2021}, but estimates of \(\theta\) would be needed
if one was interested in quantiles of the original distribution \(G\).
In the presence of clustering of extreme values, one strategy would be
to consider the same alternative hypotheses, but use sandwich-based
estimators of the covariance for the independence likelihood
\citep{Chandler.Bate:2007} coupled with Wald tests.

One could also consider sliding-block estimators of the GEV, which are
less wasteful of data and have lower mean squared error
\citep{Bucher.Zanger:2023}. Wald-based tests could be in principle
constructed even if this estimator is not based on a likelihood, but the
overlap of the blocks would mean that testing is more complex, with
standard errors having to be found using complex bootstrap schemes
\citep{Bucher.Staud:2026}.

We considered only multiples of block length (\(m_2=cm_1\) for
\(c\in \mathbb{N}\)), whereas one may wish to test for different block
lengths, say \(m=2\) versus \(m=3\). This is an inherent restriction of
our approach, although in practice it is natural to consider aggregation
of periods, e.g., monthly to yearly. Leftover observations can be
directly incorporated in the estimation procedure even if they do not
lead to full blocks. Ideas of \citet{Simpson.Northrop:2026} might be
used when dealing with missing values in the series to account for the
difference in block lengths, if the latter are missing completely at
random.

The classical extreme-value bias-variance trade-off for choosing \(m\)
implies that smaller block lengths may lead to a poor GEV approximation
(biased extrapolation), whereas longer blocks lead to smaller sample
sizes (more variability). While minimization of the mean squared error
of a single target quantity of interest (e.g., a quantile) would seem
natural at first glance, in practice the sampling distributions of their estimators are typically very skewed and normal approximations
for them can be very poor, so the use of mean square error appears questionable.  Moreover, this approach is likely to suggest using
diﬀerent block lengths for diﬀerent targets, which may be hard to sell to practitioners.

In practice, there is sample overlap when comparing blocks of length
\(m_1\) and \(m_2=cm_1\), so tests are not independent for longer
series. As our efficiency calculations suggest that taking over-long blocks may be very undesirable and the successive null hypotheses are nested, we propose
increasing \(m\) from its smallest reasonable value and stopping when the null hypothesis of max-stability is first accepted. Rejection
of max-stability at some value of \(m\) will imply rejection at any
shorter block length, so successive testing as \(m\) increases will
control the overall error rate.

\begin{acks}[Acknowledgments]

Funding in support of this work was provided by the Natural Sciences and Engineering Research Council (RGPIN-2022-05001). Calculations were performed using the \textbf{R} \citep{Rlang} programming language through computing facilities of the \href{https://laced.hec.ca/en/about-us/}{LACED} at HEC Montréal.
\end{acks}
%

\begin{supplement}%
\stitle{Reproducibility and Supporting information}
\sdescription{
Code for reproducing the figures and simulations is available from \href{https://github.com/lbelzile/block-size}{Github}. The data used in the application are found in the
\href{https://cran.r-project.org/web/packages/mev/index.html}{\texttt{mev}} \textbf{R} package \citep{mev}, which also contains the functions for implementing the tests (\texttt{test.blocksize}) and diagnostic plots
(\texttt{qqplot.blocksize}).
\begin{itemize}\tightlist
\item Section~S1 provides additional efficiency curves for different quantile levels from the GEV.
\item Section~S2 includes plots of density approximations to maxima of $m=30$ observations from Gumbel and standard normal distributions, and of their GEV parameters.
\item Section~S3 gives details on use of an alternative marginal likelihood for estimating the parameters omitting the largest observation of each block.
\item Section~S4 provides evidence of the adequate coverage of bootstrap simultaneous confidence intervals outlined in Section~\ref{sec-parametric-bootstrap}.
\item Section~S5 contains additional power curves for simulations with left-censoring and rounding, and from data from the max-domain of attraction of a GEV.
\item Section~S6 contains details on the error rate of the likelihood ratio tests.
\item Section~S7 provides an additional illustration from the max-autoregressive model and the speed of the approximation as a function of the block length.
\end{itemize}
}
\end{supplement}


\bibliographystyle{imsart-nameyear} 
\bibliography{BlockSize}       

\clearpage

\setcounter{section}{0}
\renewcommand{\thesection}{S}

\setcounter{table}{0}
\renewcommand{\thetable}{S\arabic{table}}

\setcounter{figure}{0}
\renewcommand{\thefigure}{S\arabic{figure}}

\section*{Supplementary material}\label{supplementary-material}

\subsection{Efficiencies}\label{efficiencies}

\begin{figure}[bp!]

\centering{

\includegraphics[width=\textwidth]{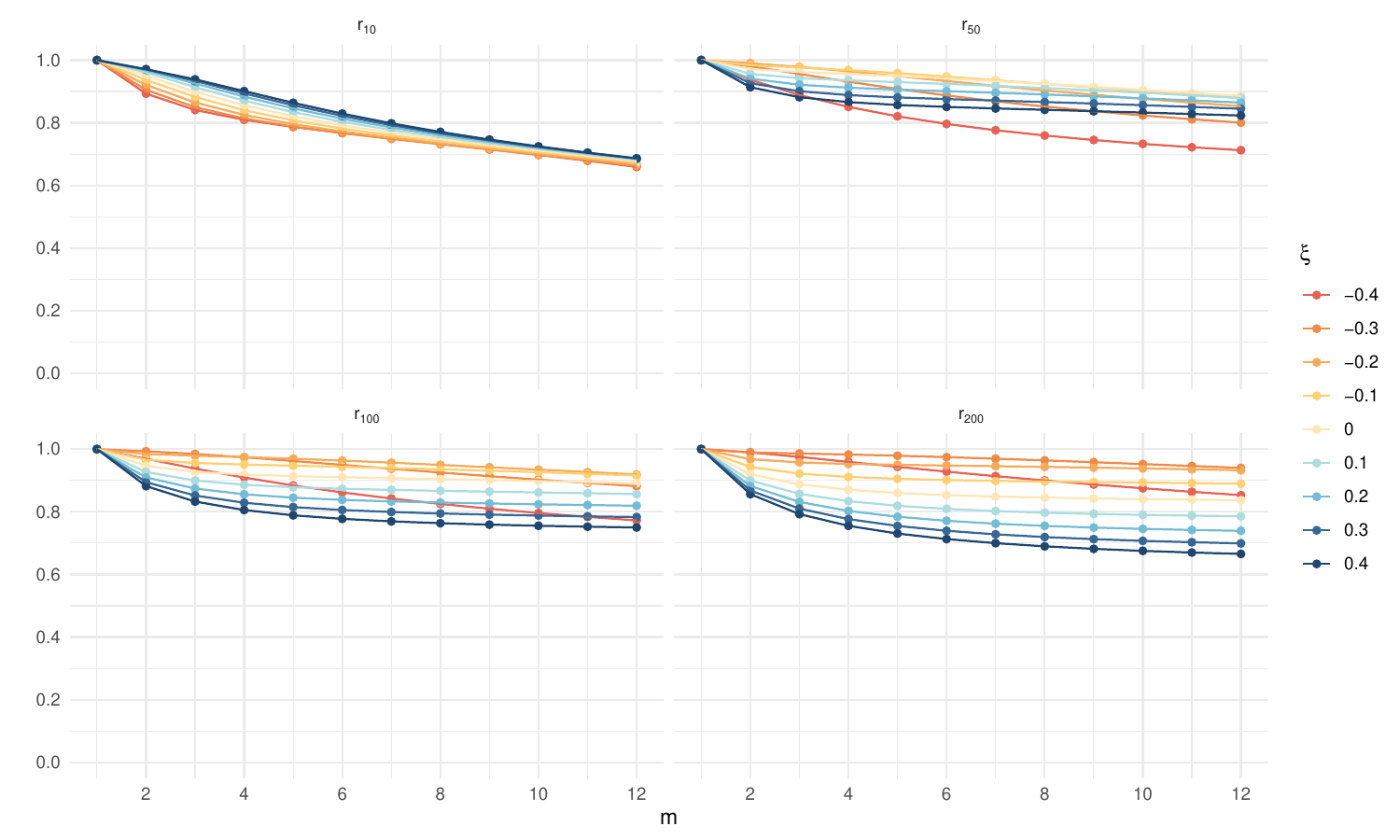}

}

\caption{\label{fig-ARE-Quantiles}Ratios of lengths of asymptotic
confidence intervals for maximum likelihood estimators of 10-, 50-, 100-
and 200-observation return levels, \(r_{10}\), \(r_{50}\), \(r_{100}\) and
\(r_{200}\), based on a random sample of \(m\) GEV observations relative
to the lengths of those based on the sample maximum only.}

\end{figure}

Figure~\ref{fig-ARE-Quantiles} shows the ratios of the lengths of
confidence intervals for $10\textrm{-}$, $50\textrm{-}$, $100\textrm{-}$ and
\(200\)-observation return levels (i.e., quantiles of the GEV at levels $0.9$, $0.98$, $0.99$ and $0.995$, based on a random sample of \(m\) GEV variates relative to
their maximum alone. The broad pattern is similar to that for the
20-year return level in Figure~\ref{fig-ARE}. The most visible change as
the return period increases is the migration of the ratios for negative
\(\xi\) from the lowest to the highest, due to the increasing efficiency
of the maximum relative to the entire sample as \(\xi\to -1/2\).

\subsection{Penultimate approximations}
\label{penultimate-approximations}

\begin{figure}

\centering{

\includegraphics[width=\textwidth]{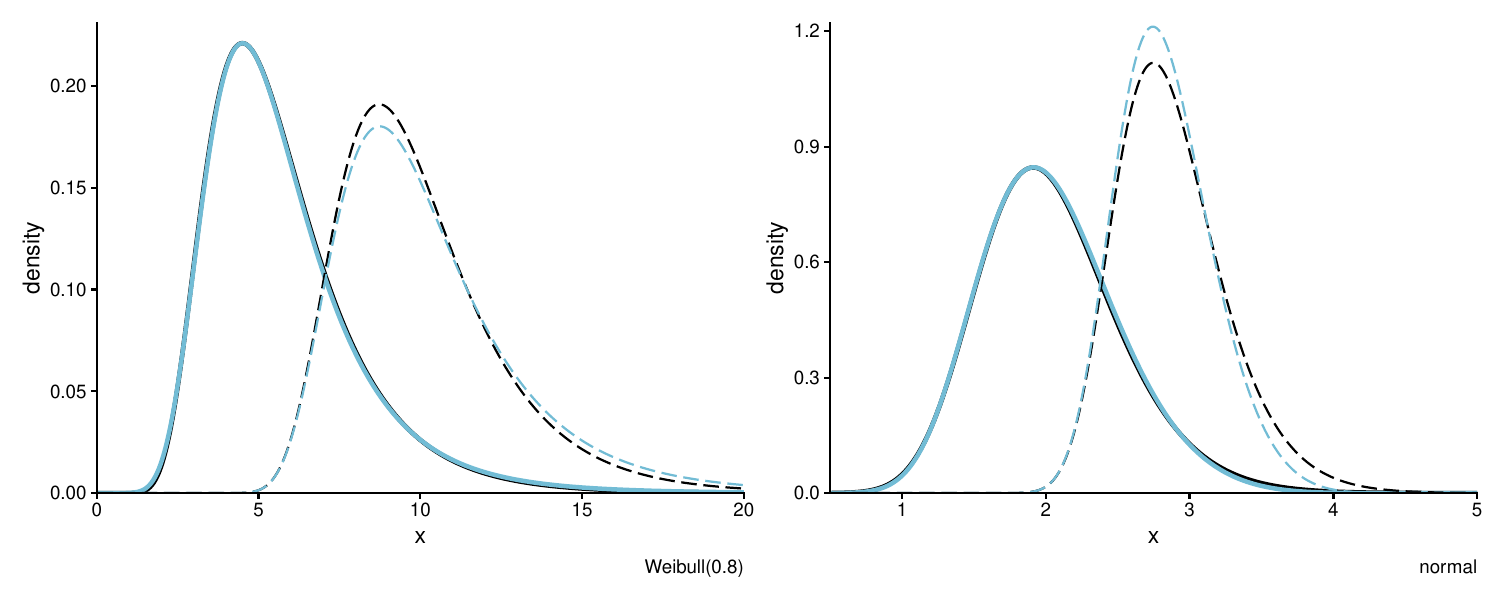}

}

\caption{\label{fig-density}Densities of GEV penultimate approximations (blue) versus true distribution \(G^{30}\) ( black) along with the corresponding extrapolation to $m=300$ using max-stability (long dash), for Weibull distribution with shape 0.8 (left) and standard normal distribution (right).}

\end{figure}

\begin{figure}

\centering{

\includegraphics[width=\textwidth]{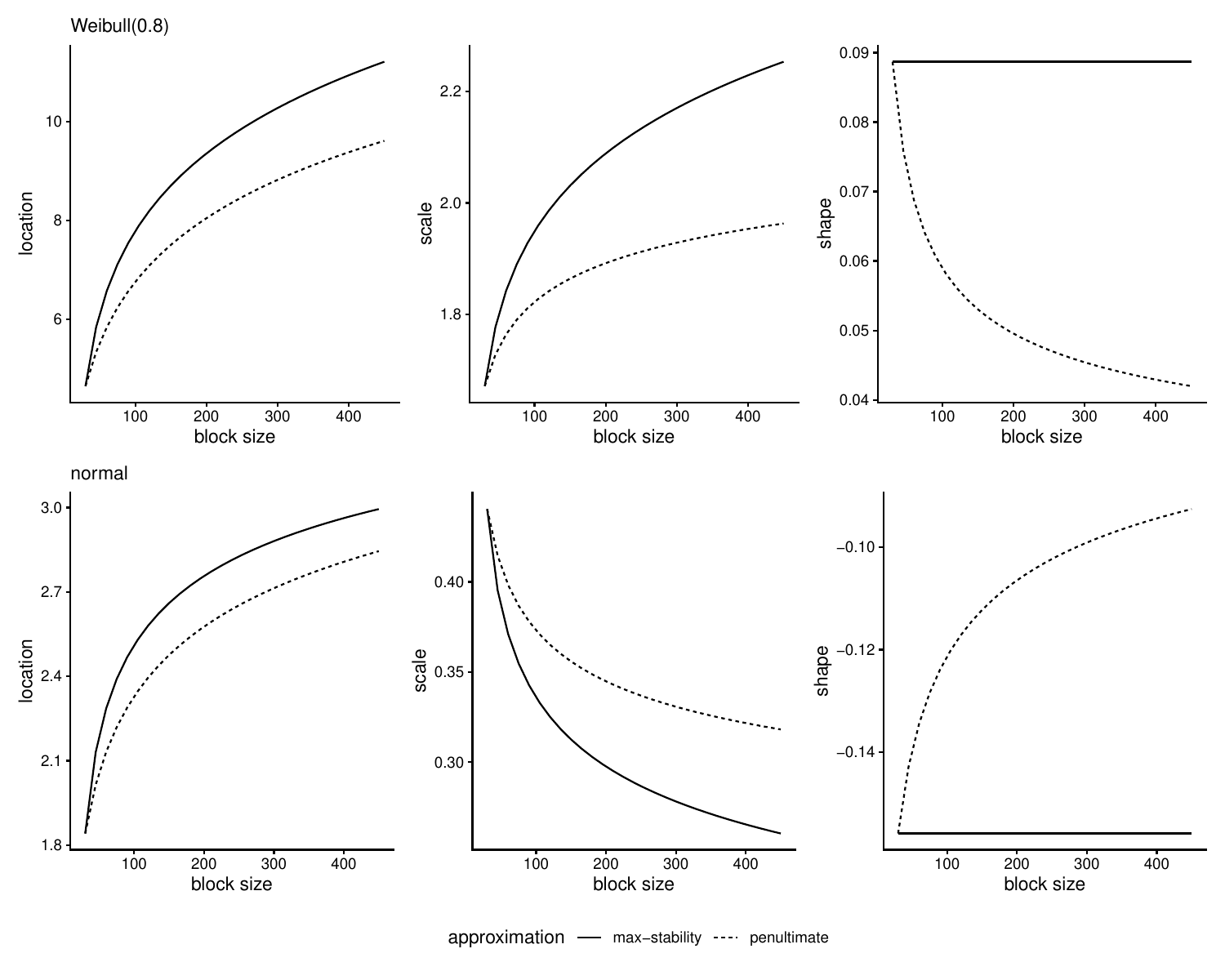}

}

\caption{\label{fig-penultnorm}Comparison between generalized extreme
value distribution parameters based on using the max-stable property
(dashed), and the penultimate approximation (full) for Weibull with shape 0.8  (top) and a standard normal
random variable (bottom). The initial
parameters extrapolated are those of the minimum block length \(m=30\).}

\end{figure}

The theory of penultimate approximation \citep[cf.][]{Smith:1987} suggests that the parameters of the best GEV fit to sample maxima should depend on the underlying distribution and  on the sample size. While the limiting shape parameter may be very different from the penultimate one, the GEV model may be already a good approximation for finite $m$. One must note, however, that small discrepancies get magnified when using the max-stability relationship to extrapolate. Figure~\ref{fig-density} shows the GEV penultimate approximation for $m=30$ and that obtained from extrapolating these parameters to maxima of $m=300$ observations by max-stability. For the latter, the discrepancy in the tails is more noticeable, even if the mode is correct. This is in contrast with the penultimate approximation for $m=300$ (not shown), which is  indistinguishable from the true distribution. Figure~\ref{fig-penultnorm} shows how the parameters of the penultimate approximation evolve when $m$ increases, as compared with the max-stability extrapolation from those with $m=30$.

Figure~\ref{fig-penultnorm} shows the penultimate approximation and the max-stability extrapolation from blocks of size \(m=30\) to
\(m=30\delta\). Both normal and Weibull distributions have limiting
shape parameter \(\xi=0\), but convergence to the limiting distribution
can be slow, as evidenced by Figure~\ref{fig-penultnorm}, even if the
GEV approximation is good for smaller block length.

\subsection{Marginal likelihood}\label{sec-marg-lik}

Sometimes it maybe of interest to use the likelihood formed from the
marginal distribution of the \((m-1)\) smallest observations of the
block, \[
\mathcal{L}_{\text{marg}}(\boldsymbol{\theta}) = \prod_{i=1}^n f_{(1),\ldots, (m-1)}(\boldsymbol{x}_i; \boldsymbol{\theta})\propto \prod_{i=1}^n\{1-F(x_{i,(m-1)})\}\prod_{j=1}^{m-1} f(x_{i,(j)};\boldsymbol{\theta})
\] which amounts to left-censoring the largest observation. Maximising
the corresponding likelihood yields the estimate
\(\widetilde{\boldsymbol{\vartheta}}\), say. We consider adding the inequality constraint
\(\sigma_0+\xi_0 (\max_{i} x_{i,(m)}-\mu_0) > 0\) to the optimization in order to ensure that the largest observation lies within the support
of the fitted GEV.

\begin{figure}

\centering{

\includegraphics[width=\textwidth]{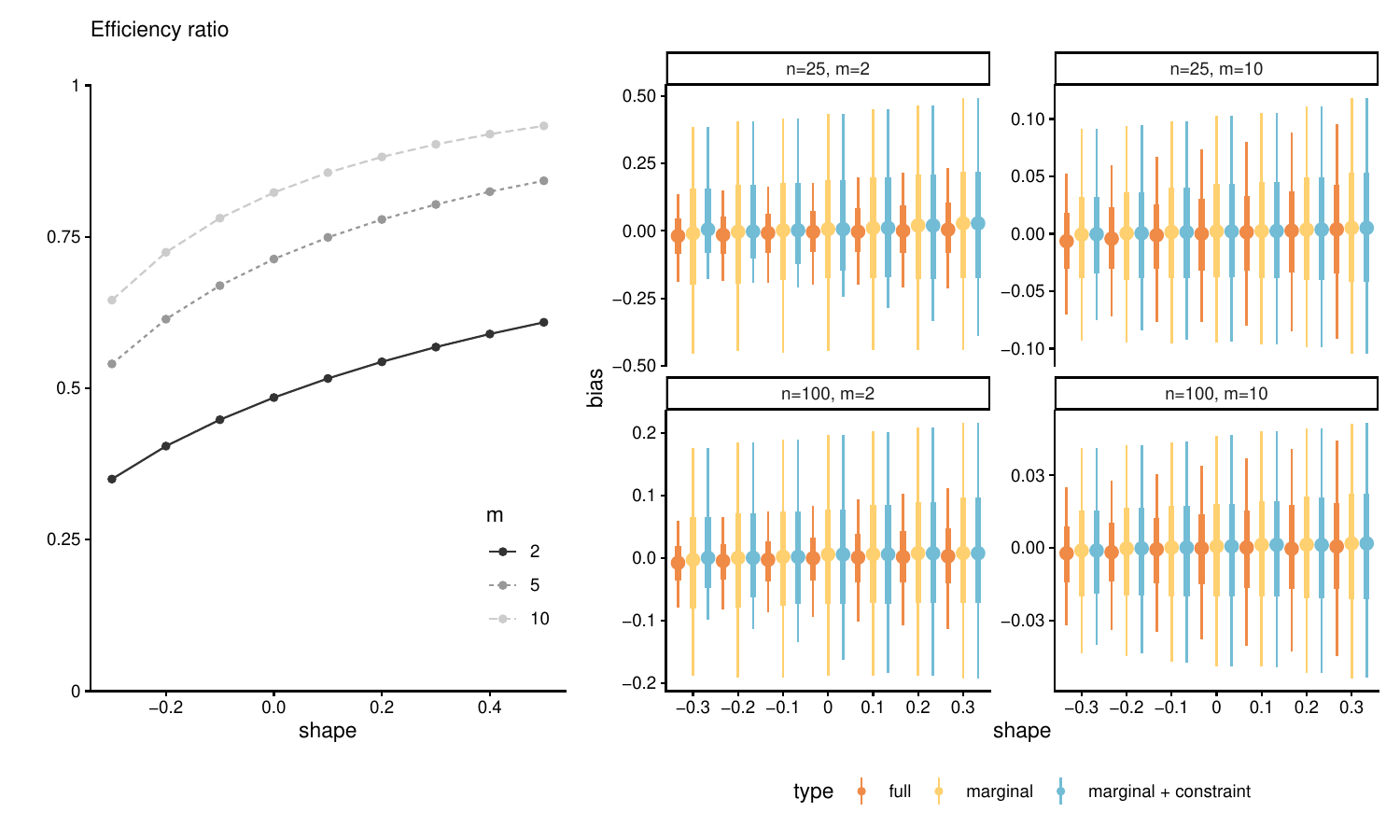}

}

\caption{\label{fig-efficiency}Comparison of marginal maximum likelihood
estimation based on \(m-1\) lowest order statistics of a sample of size
\(m \in \{2,5,10\}\) relative to full likelihood estimation, as a
function of the shape parameter \(\xi\). Efficiency curves (left) and
interval plots showing median, 50\% and 90\% intervals of the bias of
maximum likelihood estimators of the shape parameters for datasets drawn
from a GEV with shape \(\xi\) for \(m \in \{2, 10\}\) and
\(n \in \{25,100\}\), based on 1000 Monte Carlo
samples (right).}

\end{figure}

Use of the marginal likelihood avoids using the maximum both to fit the
model and to assess its fit, but leads to a loss of information: with
\(\boldsymbol{\vartheta}=(\mu=0, \sigma=1, \xi)\), the efficiency is
defined as
\(r(\boldsymbol{\vartheta})= |\mathcal{I}_{\text{marg}}(\boldsymbol{\vartheta})|^{1/3}|\mathcal{I}_{\text{full}}(\boldsymbol{\vartheta})|^{-1/3}\),
where \(\mathcal{I}\) is the Fisher information matrix. The square root of these efficiency
ratios in the left-hand panel of Figure~\ref{fig-efficiency} show more
than 50\% loss when \(m=2\), and higher loss for negative shape
parameters. These efficiencies do not reflect the
additional support constraints.

To assess the effect of the constraint on the marginal likelihood, we
conducted a small Monte Carlo experiment using datasets drawn from
\(\mathsf{GEV}(0,1, \xi)\) for different block lengths \(m\) and
replications \(n\), giving a total of \(nm\) observations, of which the
block maxima are then dropped, and the marginal likelihood based on the
others is applied. The right-hand panels of Figure~\ref{fig-efficiency}
show interval plots of biases for the shape parameters with the median
(point), along with 50\% and 90\% intervals. All estimators seem to be
approximately unbiased in large samples, but those from the marginal
likelihood are more variable. The constraint is only applied when the
shape \(\xi\) is quite negative, and the support constraint has a
reduced effect when either \(n\) or \(m\) increases.

\subsection{Coverage of simultaneous
intervals}\label{coverage-of-simultaneous-intervals}

\begin{figure}

\centering{

\includegraphics[width=\textwidth]{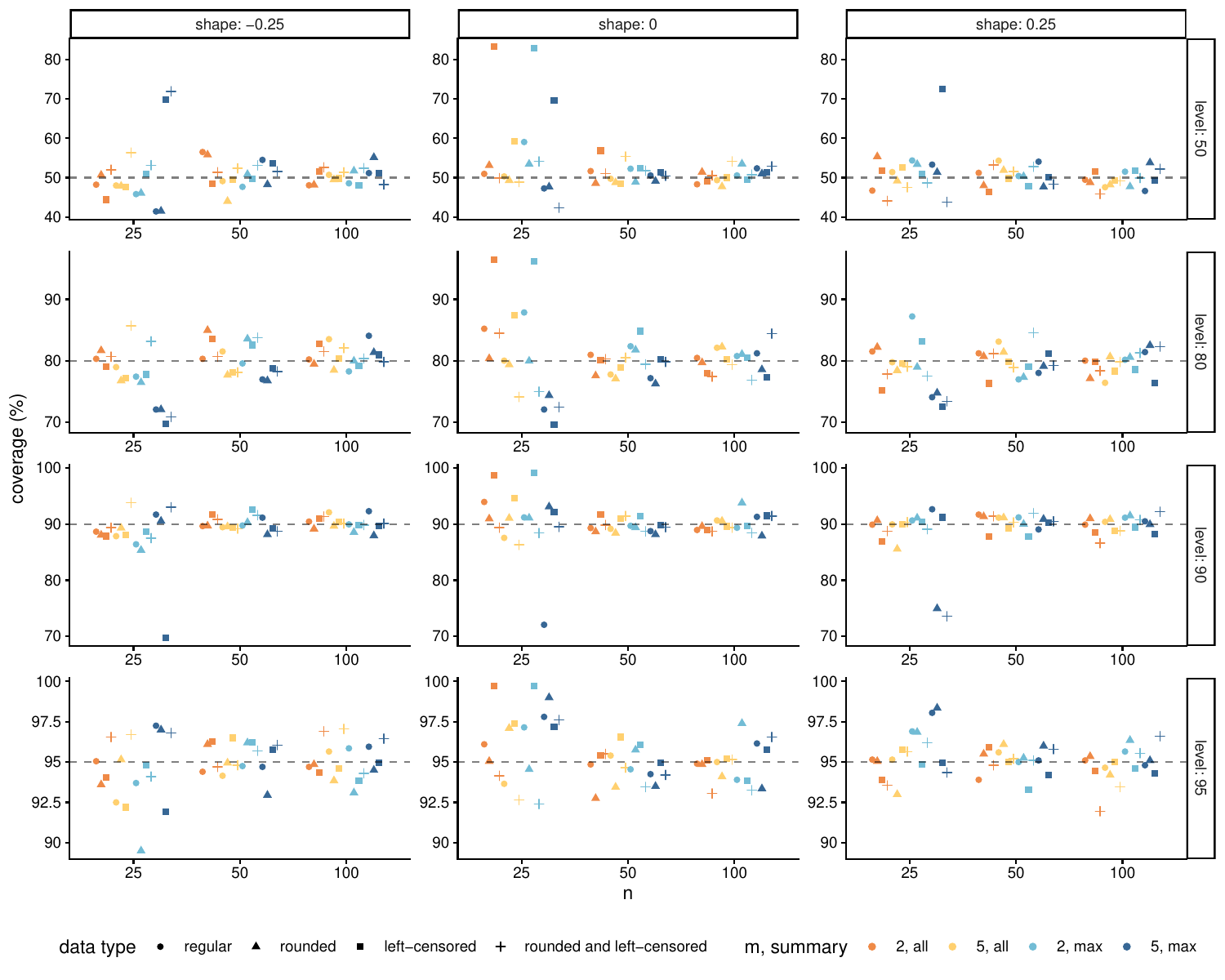}

}

\caption{\label{fig-coverage}Coverage of simultaneous confidence
intervals (in percentage) as a function of the data type, block size and
choice of summary statistic to display in probability-probability
plots.}

\end{figure}

To assess the coverage of our proposal in Section~\ref{sec-parametric-bootstrap} of the paper, we ran a simulation where we determined the
nominal rate \(\alpha^{\star}\) needed to obtain the coverage using the
bootstrap procedure; the quantiles were estimated from 500 replications.
From there, 2000 new datasets were generated with the same data type
(including rounding and left-censoring) and the coverage of 50\%, 80\%, 90\% and 95\% was calculated, for values of $m \in \{2,5\}$ and for both theset of all block maxima and the maximum $X_{(m)}$ (max). Figure~\ref{fig-coverage} shows the results: there is noticeable size distortion only when  \(n=25\), but it is non-systematic and mostly restricted to cases with left-censoring.
%

%
%
%
%
%
%

\subsection{Simulations from the maximum domain of
attraction}\label{simulations-from-the-maximum-domain-of-attraction}

To assess the impact of misspecification we simulated samples of \(n\) maxima of \(m=30\) observations from \(G\), the latter being either the
standard normal distribution function or that of a Weibull variate with
shape \(0.8\). Under the alternative we drew observations
\(Y_{1}, \ldots, Y_{m}\) from distribution \(G^{30}\), and replaced the
largest \(Y_{(m)}\) by a draw from \(G^{30\delta}\) truncated below at
\(Y_{(m-1)}\). Since the GEV approximation is excellent even with
\(m=30\), the power curves do not differ qualitatively from those
obtained for the corresponding GEV model.

\begin{figure}

\centering{

\includegraphics[width=\textwidth]{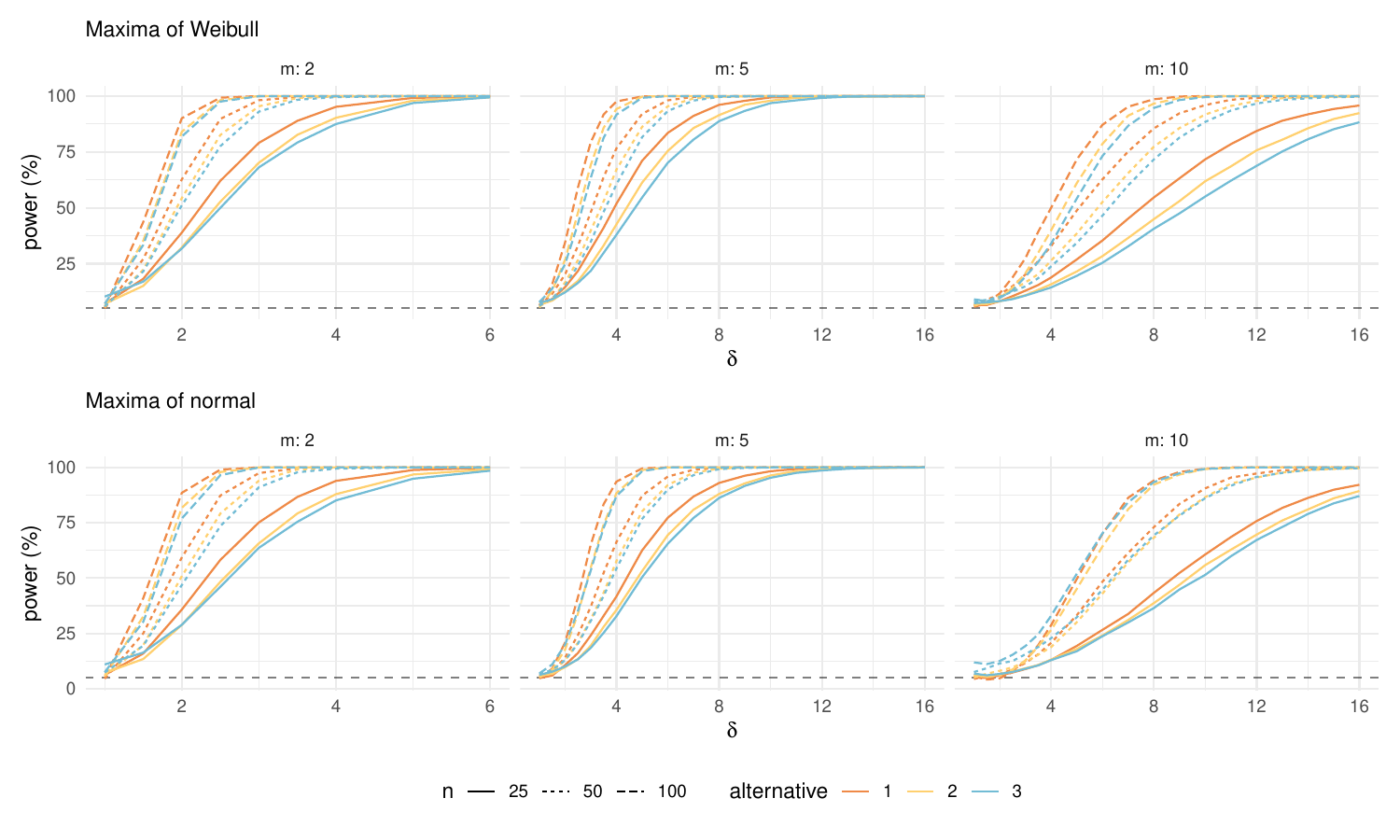}

}

\caption{\label{fig-power-curves-mda}Power curves for simulations from
\(G^{30}\), with the largest observation within block (\(m\)th) drawn
from \(G^{30\delta}\) and left-truncated at \(x_{(m-1)}\). The
distribution \(F\) is that of a Weibull distribution with shape
parameter 0.8 (top) or a standard normal distribution (bottom). The
power is calculated from samples of size \(n \times m\) observations.
Panels show the effect of changing the number of observations evaluated
(panels, from left to right), the number of additional parameters under
the alternative hypothesis considered (color) and the sample size (line
type).}

\end{figure}

The resulting power curves, in Figure~\ref{fig-power-curves-mda}, show that the test can detect departures from max-stability,
particularly if the behaviour of the maximum differs from that of other
order statistics. Power is higher for smaller \(m\), for
alternative \(A_1\), and for larger samples.

We may wonder how the curves from Figure~\ref{fig-power-curves-mda}
differ from those in the middle and bottom panel of
Figure~\ref{fig-power-curves}. The curves are nearly overlaid when
\(m=2\), but those for the max-domain of attraction for higher values of
\(m\) are slightly shifted to the left due to size distortion.

\subsection{Size of tests}\label{size-of-tests}

We computed the error rate (relative to the nominal level of 95\%) for data from the GEV (whose parameters are based on the penultimate approximation), and directly from the misspecified models $G^{30}$; see Tables~\ref{tblsizeGEV} and \ref{tblsizeMDA}. The size distortion is larger for the latter than for GEV data. The hypothesis test with alternative $A_3$ suffers from size distortion, and in general the error decreases with the sample size when the data are GEV-distributed.

\begin{table}[htbp]
  \centering
   \caption{Size of tests (\%) at level 5\% for the null hypothesis of max-stability when simulating from GEV penultimate approximations for different sample sizes $n$, distributions and alternatives (alt). Stars indicate samples for which a Kolmogorov--Smirnov test rejects at level 5\% the hypothesis that the $p$-values are uniform.}
  \label{tblsizeGEV}
  \begin{tabular}{*{2}{l}*{9}{r}}
    \toprule
     & \( m \) & \multicolumn{3}{c}{2} & \multicolumn{3}{c}{5} & \multicolumn{3}{c}{10} \\
    \cmidrule(lr){3-5} \cmidrule(lr){6-8} \cmidrule(lr){9-11}
    distribution & alt \textbar\ \( n \) & \multicolumn{1}{c}{25} & \multicolumn{1}{c}{50} & \multicolumn{1}{c}{100} & \multicolumn{1}{c}{25} & \multicolumn{1}{c}{50} & \multicolumn{1}{c}{100} & \multicolumn{1}{c}{25} & \multicolumn{1}{c}{50} & \multicolumn{1}{c}{100} \\
    \midrule
    Weibull & $A_1$ & $\star$ 6.3 & 5.7 & 5.7 & 6.0 & 5.9 & 5.1 & 5.5 & 5.5 & 5.1 \\
    & $A_2$ & $\star$ 6.0 & $\star$ 6.1 & 5.8 & $\star$ 6.7 & 5.3 & 4.5 & $\star$ 6.0 & 5.5 & $\star$ 5.5 \\
    & $A_3$ & $\star$ 7.9 & $\star$ 7.9 & $\star$ 6.9 & $\star$ 7.8 & 6.2 & 5.5 & $\star$ 6.6 & 5.5 & $\star$ 6.3 \\ \addlinespace[3pt]
    normal & $A_1$ & $\star$ 5.8 & 5.2 & 5.5 & 5.8 & 5.6 & 5.1 & 5.1 & 5.8 & 5.0 \\
    & $A_2$ & $\star$ 5.7 & $\star$ 6.0 & $\star$ 5.6 & 5.9 & 5.8 & 4.8 & 4.4 & 5.8 & $\star$ 5.1 \\
    & $A_3$ & $\star$ 8.8 & $\star$ 8.4 & $\star$ 7.0 & $\star$ 7.0 & $\star$ 6.4 & 5.8 & $\star$ 5.1 & 5.8 & $\star$ 5.6 \\
    \bottomrule
  \end{tabular}
\end{table}

\begin{table}[htbp]
\centering
  \caption{Size of tests (in percentage) at level 5\% for the null hypothesis of max-stability when simulating from the max-domain of attraction for different distributions and alternatives (alt). Stars indicate samples for which a Kolmogorov--Smirnov test of uniformity rejects the null hypothesis at level 5\%.}
    \label{tblsizeMDA}
\begin{tabular}{*{2}{l}*{9}{r}}
    \toprule
     & \( m \) & \multicolumn{3}{c}{2} & \multicolumn{3}{c}{5} & \multicolumn{3}{c}{10} \\
    \cmidrule(lr){3-5} \cmidrule(lr){6-8} \cmidrule(lr){9-11}
    distribution & alt \textbar\ \( n \) & \multicolumn{1}{c}{25} & \multicolumn{1}{c}{50} & \multicolumn{1}{c}{100} & \multicolumn{1}{c}{25} & \multicolumn{1}{c}{50} & \multicolumn{1}{c}{100} & \multicolumn{1}{c}{25} & \multicolumn{1}{c}{50} & \multicolumn{1}{c}{100} \\
    \midrule
    Weibull & $A_1$ & $\star$ 6.9 & 5.2 & 5.0 & $\star$ 5.8 & 5.5 & $\star$ 6.0 & $\star$ 6.2 & 6.9 & 5.7 \\
    & $A_2$ & $\star$ 7.1 & 4.5 & 5.4 & $\star$ 6.3 & 5.3 & $\star$ 5.3 & $\star$ 7.0 & $\star$ 7.2 & $\star$ 7.2 \\
    & $A_3$ & $\star$ 9.6 & $\star$ 5.9 & $\star$ 6.7 & $\star$ 8.4 & $\star$ 7.0 & $\star$ 8.6 & $\star$ 7.3 & $\star$ 9.0 & $\star$ 10.1 \\ \addlinespace[3pt]
    normal & $A_1$ & $\star$ 6.2 & 4.7 & 4.9 & 4.9 & 4.9 & 4.9 & 5.7 & 5.3 & 5.9 \\
    & $A_2$ & $\star$ 6.5 & 4.9 & $\star$ 5.6 & $\star$ 5.6 & 5.4 & $\star$ 5.7 & $\star$ 5.4 & $\star$ 6.3 & $\star$ 6.9 \\
    & $A_3$ & $\star$ 10.0 & $\star$ 6.6 & $\star$ 5.9 & $\star$ 6.4 & $\star$ 6.1 & $\star$ 7.5 & $\star$ 6.5 & $\star$ 8.7 & $\star$ 12.0 \\
    \bottomrule
  \end{tabular}

\end{table}

\subsection{First-order max-autoregressive
process}\label{first-order-max-autoregressive-process}

\begin{figure}

\centering{

\includegraphics[width=\textwidth]{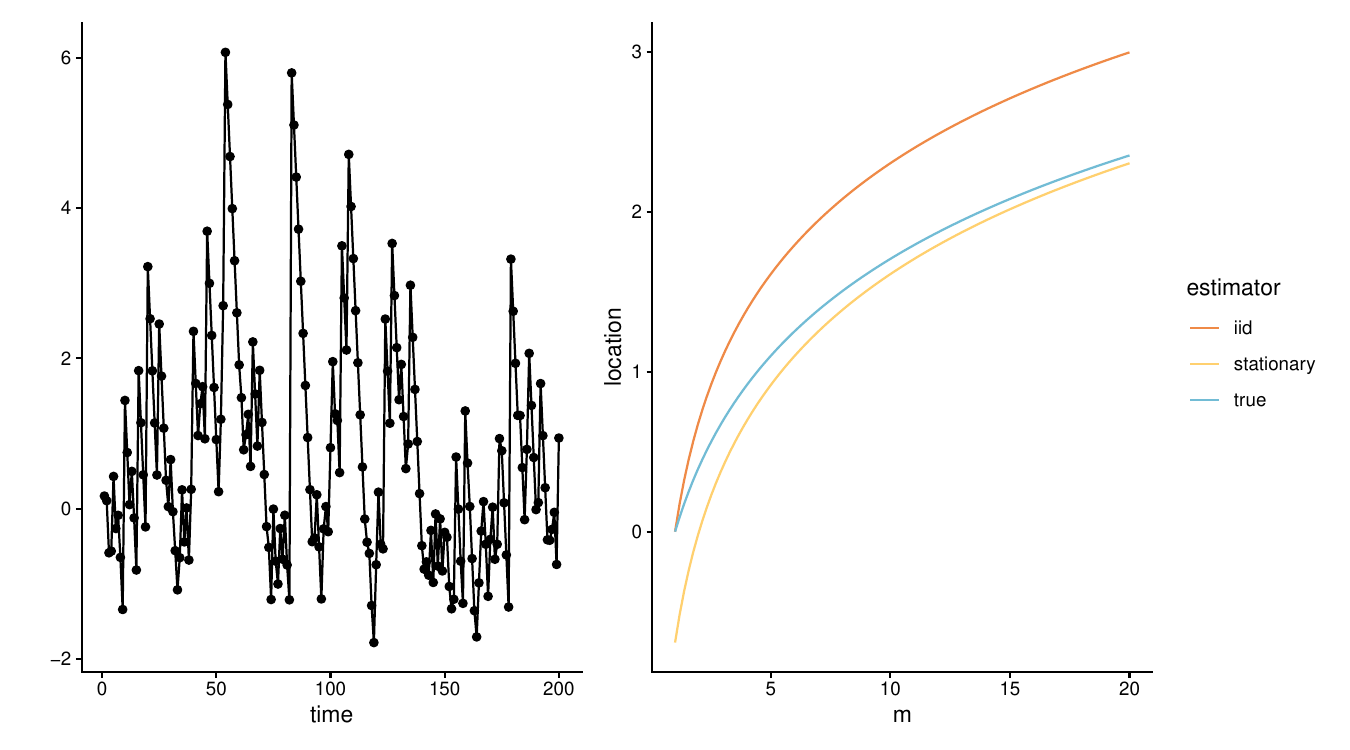}

}

\caption{Max-autoregressive model with \(\xi=0\)
and \(\theta=0.5\). Simulated time series of length $n=200$ (left), and
location parameters of the Gumbel maxima based on extrapolating to
blocks of length \(m\) by taking \(\log(m)\) (iid), and
\(\log(m\theta)\) (stationary) as location parameters relative to the
true distribution (right).}
\label{fig-tavares77}

\end{figure}

Figure~\ref{fig-tavares77} shows a sample of observations drawn from the max-autoregressive model with \(\theta=0.5\) and \(\xi=0\), corresponding to clusters
of average size two (left panel). Block maxima are also Gumbel
distributed, and only the location parameter changes with \(m\) due to
max-stability. The resulting curves, obtained from extrapolating from
\(m=1\), are shown in the right-hand panel of
Figure~\ref{fig-tavares77}: using the limiting value leads to slight
underestimation of \(\mu\) and an incorrect assumption of independence
leads to overestimation of \(\mu\).

\end{document}